\documentclass[12pt]{article}
\usepackage{amsmath,amssymb}
\usepackage{amsfonts,mathrsfs}
\usepackage{graphicx,epsfig}
\usepackage[hypertex]{hyperref}
\usepackage{twist_nlie}

\numberwithin{equation}{section}

\begin{document}

{\ }
\vspace{-10mm}

\begin{flushright}
{\bf February 2012}\\[1mm]
{\small ITP-UU-12/06}\\[1mm]
{\small SPIN-12/05}\\[1mm]
\end{flushright}

\vskip 2cm

\begin{center}
\LARGE

\mbox{\bf Contour deformation trick in hybrid NLIE}


\vskip 2cm
\renewcommand{\thefootnote}{$\alph{footnote}$}

\large
\centerline{\sc Ryo Suzuki\footnotetext{{\tt R.Suzuki@uu.nl}}}

\vskip 1cm

\emph{Institute for Theoretical Physics and Spinoza Institute,\\
Utrecht University, 3508 TD Utrecht, The Netherlands}

\end{center}

\vskip 14mm

\centerline{\bf Abstract}

\vskip 6mm

The hybrid NLIE of \AdSxS\ is applied to a wider class of states.
We find that the Konishi state of the orbifold \AdSxSZ\ satisfies $A_1$ NLIE with the source terms which are derived from contour deformation trick.
For general states, we construct a deformed contour with which the contour deformation trick yields the correct source terms.

\vskip 2mm

\vfill
\thispagestyle{empty}
\setcounter{page}{0}
\setcounter{footnote}{0}
\renewcommand{\thefootnote}{\arabic{footnote}}
\newpage

\tableofcontents

\newpage

\section{Introduction and Summary}\label{sec:Intro}

The primary example of AdS/CFT correspondence is the one between four-dimensional $\cN=4$ super Yang-Mills and \AdSxS\ string theory \cite{Maldacena97}.
The spectrum of string states on \AdSxS\ can be computed by the mirror Thermodynamic Bethe Ansatz (TBA) equations \cite{BFT09, GKKV09, AF09b} based on string hypothesis in the mirror model \cite{AF07, AF09a}; or equivalently the extended Y-system on $\alg{psu}(2,2|4)$-hook \cite{GKV09a,CFT10,BH11a}.
It is believed that these methods give the exact answer, because they capture all finite-size corrections \cite{AJK05, BJ08}.

The numerical study of the mirror TBA has made progress \cite{GKV09b, Frolov10, LM11, Frolov12a}. However, it suffers from the problem of critical coupling constants \cite{AFSu09}.
The analyticity of the unknown variables called Y-functions changes around certain values of 't Hooft coupling constant, and the explicit form of the TBA equations changes there discontinuously.
As a result, it is difficult to solve the equation with high precision around the critical values, and to judge if the exact energy does not show unusual behavior like inflection points around the critical points.

The author has recently applied the method of hybrid nonlinear integral equations (hybrid NLIE) \cite{Su99} to the mirror TBA for \AdSxS\ \cite{Suzuki11}.
This method replaces the horizontal part of the mirror TBA equations by $A_1$ NLIE.\footnote{This equation is called Kl\"umper-Batchelor-Pearce or Destri-de Vega equation in the literature \cite{KB90,KBP91,DdV92a,DdV92b,DdV94}. We call it $A_1$ NLIE, since it can be derived from $A_1$ TQ-relations and analyticity conditions as shown in \cite{Suzuki11}.}
The hybrid NLIE consists of a smaller set of unknown variables than the mirror TBA, and we expect that it suffers less often from the problem of critical coupling constants. We exemplify our expectation in a way similar to \cite{AFSu09}.

For this purpose the mirror TBA for the twisted \AdSxS\ offers a desired playground, because all Y-functions have intricate analytic properties, depending on the twist angle $\alpha$ and 't Hooft coupling constant $g = \frac{\sqrt \lambda}{2\pi}$.\footnote{In fact, the mirror TBA for $Y_{M|w}$ in the untwisted model do not have critical coupling constants asymptotically. We checked this claim for several four particle states for $g \lesssim 1$.}
The orbifold Konishi state is the simplest nontrivial example that exhibits critical behavior in the mirror TBA for $Y_{M|w}$. For this state, we find that the hybrid NLIE also exhibits critical behavior; its source terms change discontinuously across certain values of coupling constant.

Orbifold Konishi is a two-particle state in the $\sl(2)$ sector of \AdSxSZ, where the $\bb{Z}_S$ acts on $\alg{su}(2)^2 \subset [\alg{su}(2|2)^2 \cap \alg{su}(4)]$. This is also a special state in the twisted \AdSxS, $\beta$- or $\gamma$-deformed \AdSxS\ models. The orbifold and $\gamma$-deformed models are another important examples of AdS/CFT correspondence, realized in gauge theory \cite{Roiban03,BC04,Ideguchi04,BR05a,BR05b,AKS06,AKS07} and in string theory \cite{LLM04,LM05,Frolov05,FRT05a,FRT05b}.
Finite-size corrections of deformed theories have been studied in gauge theory \cite{FSSZ08b,FSSZ08c,MS11}, in string theory \cite{BF08}, by L\"uscher formula \cite{Gunnesson09,BdA09,BHJL09,dLL10,BLMM10,ABBN10a,ABBN11,ABK12}, and by the mirror TBA or Y-system \cite{GL10,AdLvT10,dLvT11,dLvT12a,BM11a}.
However, it is not clear if the corresponding sigma model on twisted \AdSxS\ possesses integrability (see \cite{Zoubos10} for review), though integrable twists exist mathematically.

Next, we notice that such discontinuous change of the NLIE for the orbifold Konishi state can be explained by the contour deformation trick.
There is a conjecture that the TBA for excited states follows from the TBA for the ground state by analytic continuation of coupling constant \cite{DT96,DT97}.
It is expected that such analytic continuation introduces extra singularities of the integrand on the complex rapidity plane, and deforms the integration contour accordingly.
Then, the excited states TBA should be expressed equivalently either as the ground-state TBA integrated over the deformed contour, or as the TBA integrated over the real line with additional source terms.
This idea is called contour deformation trick.
The contour deformation trick predicts how to correct the TBA when numerical iteration ceases to converge due to the change of analyticity, and is a guideline to study various states in the mirror TBA \cite{AFSu09} including boundstates \cite{AFvT11}.
The $A_1$ NLIE with source terms has been studied in various examples \cite{FMQR96,Hegedus03,Su04,Hegedus04,Hegedus05,HRS06,Hegedus07,GKKV08,Caetano10}, and the contour deformation trick was used in \cite{KP92,Klumper93}.

With successful examples of the contour deformation in mind, we ask what the most general possible source terms are, and if they are obtained by the contour deformation trick. In principle, $A_1$ NLIE can be derived even when the Q-functions are meromorphic, rather than analytic, in the upper or lower half plane. Then the isolated singularities of Q-functions provide extra source terms to $A_1$ NLIE.
It is a nontrivial question whether such source terms can be explained by the contour deformation trick, particularly with the same contour as in the orbifold Konishi state.
Indeed, mismatch is found between the two results. To reconcile this problem, we construct a deformed contour which is consistent for general states including orbifold Konishi.
The consistent deformed contour picks up only the preferred singularities of the integrand and runs both the lower and upper half planes.
The details will be discussed in Section 3.

The contour deformation trick illustrates the difference between hybrid NLIE and FiNLIE \cite{GKLV11}.
In the latter the integrals run over the gap discontinuity of dynamical variables, which is not something to be deformed. 
In contrast, hybrid NLIE is written in terms of gauge-invariant (but frame-dependent) variables,\footnote{See the discussion at the end of Appendix \ref{app:twisteded Wronskian} for the frame dependence.} allowing us to handle the equations similar to that of the mirror TBA.

\bigskip
This paper is organized as follows.
In Section 2, we study the orbifold Konishi state from the mirror TBA and hybrid NLIE, and clarify the critical behavior in the asymptotic limit.
In Section 3, we discuss the source terms of $A_1$ NLIE in view of contour deformation trick.
Section 4 is for conclusion.
In appendices, we introduce our notation, review the NLIE variables, compute the asymptotic transfer matrix in the form of Wronskian, and derive the results in Section 3.

\section{TBA and NLIE for twisted \AdSxS}\label{sec:twisted AdSxS}

We study the critical behavior of hybrid NLIE for the orbifold Konishi state as a specific example. We briefly review the mirror TBA in twisted \AdSxS\ and their critical behavior.

\subsection{Orbifold Konishi state}\label{sec:orbKon}

The orbifold Konishi state can be defined in two equivalent ways.

The first is to consider the $\sl(2)$ Konishi descendant on the orbifold \AdSxSZ, where the $\bb{Z}_S$ action is chosen as follows (see \cite{AdLvT10}). We decompose the transverse 8+8 fields of \AdSxS\ into $({\bf 2}|{\bf 2}) \otimes ({\bf 2}|{\bf 2})$ representation of $\alg{su}(2|2)_L \times \alg{su}(2|2)_R$\,, as
\begin{equation}
\( \Phi^I \,, D_\mu Z \,, \Psi \,, \overline{\Psi} \) \quad \leftrightarrow \quad
\( Y_{b \dot b} \,, Y_{\beta \dot \beta} \,, Y_{b \dot \beta} \,, Y_{\beta \dot b} \)
\equiv (y_b \, y_{\dot b} \,, \eta_\beta \, \eta_{\dot \beta} \,, y_b \, \eta_{\dot \beta} \,, \eta_\beta \, y_{\dot b} ),
\label{transverse fields}
\end{equation}
where $b, \dot b = 1,2$ refer to the ${\rm S}^5$ part, and $\beta, \dot \beta = 3,4$ refer to the ${\rm AdS}_5$ part of $\alg{su}(2|2)^2$. The boundary conditions of $y_b$ are twisted by $\bb{Z}_S$ as
\begin{equation}
\begin{pmatrix}
y_1 (\sigma = 2\pi) \\
y_2 (\sigma = 2\pi)
\end{pmatrix}
=
\begin{pmatrix}
e^{+i \alpha_L} & 0 \\
0 & e^{-i \alpha_L}
\end{pmatrix}
\begin{pmatrix}
y_1 (\sigma = 0) \\
y_2 (\sigma = 0)
\end{pmatrix}, \qquad
\alpha_L = \frac{2 \pi n_L}{S} \quad (n_L \in \bb{Z}).
\label{def:ZS action}
\end{equation}
Similarly, the boundary conditions of $y_{\dot b}$ are twisted by $\alpha_R = \frac{2 \pi n_R}{S}$\,.
The orbifold action \eqref{def:ZS action} affects only the auxiliary part of the asymptotic Bethe Ansatz equations.
Thus, if we set the total momentum to zero as in the ordinary Konishi state, the asymptotic Bethe roots remain unchanged before and after orbifolding. This is called orbifold Konishi state.

The second is to introduce integrable twisted boundary conditions to the transfer matrix of \AdSxS.
To preserve the integrability, the twist operator must commute with the S-matrix.
When the twist operator belongs to $[ \alg{su}(2|2)^2 \cap \alg{su}(4) ]$ and the twist angle is equal to a multiple of $2\pi/S$, Konishi state of the twisted \AdSxS\ is equivalent to the orbifold Konishi state.

The second point of view is useful to construct the twisted transfer matrix, as defined by
\begin{equation}
T_{Q,1}^L = {\rm str}_Q \[ g_0 \, \bb{S}_{01} \, \bb{S}_{02} \, \dots \, \bb{S}_{0N} \], \qquad
g_0 = {\rm diag} \( e^{+i\alpha_L} \,, e^{-i\alpha_L} \,, 1, 1\).
\label{def:twisted transfer L}
\end{equation}
and similarly for $T_{Q,1}^R$. The $\bb{S}_{0i}$ is the S-matrix between the mirror particle and the $i$-th particle in string theory. We can diagonalize \eqref{def:twisted transfer L} by algebraic Bethe Ansatz \cite{ABBN10b}. In practice, it is easier to twist the generating function for the eigenvalues of transfer matrices \cite{Beisert06b, AdLvT10, GL10}.
This construction will be discussed in Appendix \ref{app:twisteded Wronskian}, where we also rewrite the transfer matrices in the form of Wronskian.
In what follows we set $\alpha_L = \alpha_R \equiv \alpha$ for simplicity.

The mirror TBA for the twisted model is obtained as follows.
The twist angle $\alpha$ in string theory corresponds to the insertion of defect operator in mirror theory \cite{ABBN11}. In particular, the same mirror string hypothesis is used in both twisted and untwisted models. In the case of orbifold, the defect operator can be identified as an extra chemical potential, and it changes the $v \to \pm\infty$ asymptotics of Y-functions \cite{dLvT11}.
The mirror TBA equations for twisted \AdSxS\ are solved by the twisted transfer matrices in the asymptotic limit \cite{AdLvT10}.

\subsection{TBA and NLIE in horizontal strips}

We compare mirror TBA and hybrid NLIE in the horizontal part of the $\alg{psu}(2,2|4)$-hook for the twisted \AdSxS.
We will consider only the states which are invariant under the interchange $(a,s) \to (a,-s)$ of the $\alg{psu}(2,2|4)$-hook.

The simplified TBA equation for $Y_{1|w}$ and $Y_{M|w} \ (M \ge 2)$ can be written as
\begin{align}
\log Y_{1|w} &= - V_{1|w} + \log (1+Y_{2|w}) \star s_K
+ \log \frac{1-\frac{1}{Y_-} }{1-\frac{1}{Y_+} } \hstar s_K \,,
\label{TBA Y1w} \\[1mm]
\log Y_{M|w} &= - V_{M|w} + \log (1+Y_{M+1|w}) \star s_K
+ \log (1+Y_{M-1|w}) \star s_K \,,
\label{TBA YMw}
\end{align}
where $V_{M|w}$ is the source term, which depends on the state and the values of $(\alpha, g)$ under consideration.
In the hybrid NLIE, the $1+Y_{M+1|w}$ on the right hand side is replaced by
\begin{equation*}
1+Y_{2|w} = (1+\fa_3^{\nu \, [+\gamma]} ) \, (1+\fba_3^{\nu \, [-\gamma]}), \qquad
1+Y_{M+1|w} = (1+\fa_{M+2}^{\nu \, [+\gamma]} ) \, (1+\fba_{M+2}^{\nu \, [-\gamma]}).
\end{equation*}
The pair of parameters $\pare{\fa_s^\nu \,, \fba_s^\nu} (s \ge 3)$ are determined by $A_1$ NLIE,
\begin{align}
\log \fa_s^\nu &= -J_s^\nu + \log (1+\fa_s^\nu) \star K_f
- \log (1+\fba_s^\nu) \star K_f^{[+2-2\gamma]}
+ \log (1+Y_{s-2|w} ) \star s_K^{[-\gamma]} \,,
\label{fa_s NLIE} \\[2mm]
\log \fba_s^\nu &= -\olJ_s^\nu + \log (1+\fba_s^\nu) \star K_f 
- \log (1+\fa_s^\nu) \star K_f^{[-2+2\gamma]}
+ \log (1+Y_{s-2|w} ) \star s_K^{[+\gamma]} \,,
\label{fba_s NLIE}
\end{align}
where $\nu = {\rm I}$ or ${\rm II}$ refers to the two sets of Q-functions \cite{Suzuki11}, and $\gamma\ (0 < \gamma < 1)$ is a regularization parameter, as reviewed in appendix \ref{sec:NLIE var}.
We leave $s \in \bb{Z}_{\ge 3}$ unspecified, though one can substitute $s=3$ at any time.
The case of $\nu = {\rm I}$ is simpler than $\nu ={\rm II}$, because the source terms $\{J_3^{\rm I} \,, \olJ_3^{\rm I} \}$ vanishes in the Konishi state of the untwisted \AdSxS\ model, at least asymptotically.
Below we consider the case $\fa_s^{\rm I} \,, \fba_s^{\rm I}$ only, and omit $\nu = {\rm I}$.
In short, the $Y_{M|w}$ functions of the mirror TBA are replaced by three dynamical variables, $(\fa_3 \,, \fba_3 \,, Y_{1|w})$.\footnote{If we consider both left and right horizontal strips of the $\alg{psu}(2,2|4)$-hook, $Y_{M|w}^{(L)} \,, Y_{M|w}^{(R)}$ are replaced by six dynamical variables, $(\fa_3^{(L)} \,, \fba_3^{(L)} \,, Y_{1|w}^{(L)} \,, \fa_3^{(R)} \,, \fba_3^{(R)} \,, Y_{1|w}^{(R)})$.}

Numerically, the equations \eqref{fa_s NLIE}, \eqref{fba_s NLIE} can be checked modulo multiple of $\pi i$ for the following reason. Since $(1+\fa_s \,, 1+\fba_s)$ are complex, their logarithm may choose either of $\log (-1) = \pm \pi i$, which changes the numerical value of the convolution by $(2 \pi i) \star K_f = \pi i$.

\paragraph{Critical lines and analyticity.}

The source terms in TBA or NLIE change discontinuously as we vary the parameters $(\alpha , g)$.
We divide the $(\alpha , g)$ plane into subregions according to different form of the source terms.
The boundary of subregions is called critical lines.
We denote the critical lines by $\alpha = \alpha_{\rm cr}^{(i)} (g)$ or $g = g_{\rm cr}^{(i)} (\alpha)$.

The critical lines are different for different integral equations of TBA or NLIE.
So the phase space of a given state in the twisted \AdSxS\ is divided into infinitely many tiny regions as
\begin{alignat}{9}
g_{\rm cr}^{(I)} (\alpha) &= \Bigl\{ \bigcup_{(a,s) \in {\rm T-hook}} g_{\rm cr}^{(i)} (\alpha) [Y_{a,s}] \Bigr\}
&\qquad &\text{for TBA},
\\[1mm]
g_{\rm cr}^{(I)} (\alpha) &= \Bigl\{ \bigcup_{(a,|s| \le 2) \in {\rm T-hook}} g_{\rm cr}^{(i)} (\alpha) [Y_{a,s}] \Bigr\}
\cup \Bigl\{ g_{\rm cr}^{(j)} (\alpha) [\fa_3 \,, \fba_3] \Bigr\}
&\qquad &\text{for hybrid NLIE}.
\end{alignat}

The critical lines, or discontinuous changes of source terms, come from the change of the analyticity of unknown variables in a given integral equation. This statement holds true for both simplified TBA and NLIE.
The TBA for the orbifold Konishi has already been studied in detail \cite{dLvT11}, so we will make this statement more precise for the NLIE.

It should be noted that the critical lines of hybrid NLIE depend on the regularization parameter $\gamma$.
Also, the critical lines of the mirror TBA change if we pull back the deformed contour to the line $\bb{R}+i \delta$ with $\delta \neq 0$ instead of the real line.\footnote{It is not practical to solve the mirror TBA using the Y-functions not sitting on the real axis, because the reality of Y-functions is abandoned.}
Besides its simplicity, there is no particular meaning of setting $\gamma$ or $\delta$ to zero.
From the continuity of the equations this implies that physical quantities such as the exact energy should not be singular at $g = g_{\rm cr}^{(l)} (\alpha)$\,.\footnote{The author thanks a referee of JHEP for pointing this out.}

\subsection{Source terms of $A_1$ NLIE}\label{sec:src A1 NLIE}

We determine the source terms in $A_1$ NLIE $(J_s \,, \olJ_s)$, by taking examples of the twisted ground state and orbifold Konishi state.

\paragraph{Source term of twisted ground state.}

The ground state of the twisted \AdSxS\ satisfies the simplified mirror TBA with $V_{M|w} = 0$ \cite{dLvT11}. It also satisfies the hybrid NLIE with the chemical potential
\begin{equation}
J_s = +i \alpha, \qquad
\olJ_s = -i \alpha .
\label{ground state Js}
\end{equation}
This result follows immediately from the asymptotic solution discussed in Appendix \ref{app:twisteded Wronskian}.
Even for excited states, each term in the $A_1$ NLIE approaches its ground state value in the limit $v \to \pm \infty$, just like TBA.
Furthermore, the orbifold Konishi state satisfies the same equation at small $\alpha \neq 0$ and small $g$.
For general $(\alpha, g)$ we should add logarithms of S-matrix to the source term.

\paragraph{Main strip of hybrid NLIE.}

Before studying source terms at general $(\alpha, g)$, let us discuss the main strip of the mirror TBA or the hybrid NLIE. The main strip is defined by the region of complex plane in which the respective equation remains valid without modification.
It is helpful to identify the main strip in advance, because the critical lines are often related to the movement of extra zeroes going in or out of this strip.

The main strip of the simplified TBA for $Y_{M|w}$ \eqref{TBA Y1w}, \eqref{TBA YMw} is $\cA_{-1,1}$ defined in \eqref{def:cA}. This is because we encounter the singularity of $s_K$ along the boundary of $\cA_{-1,1}$\,. 
Analytic continuation of the simplified TBA beyond $\cA_{-1,1}$ requires us to add an extra term $\sim \log (1+Y^\pm)$ for some $Y$.

The main strip for the hybrid NLIE is smaller than that of the simplified TBA. Consider the holomorphic part of $A_1$ NLIE \eqref{fa_s NLIE}, which contains the kernels $K_f \,, K_f^{[+2-2\gamma]}\,, s_K^{[-\gamma]}$\,. Since these kernels are singular at $K_f (\pm 2i/g)$ and $s_K (\pm i/g)$, the main strip of \eqref{fa_s NLIE} is
\begin{equation}
{\rm Im} \, v \in \(- \frac{1-\gamma}{g} \,, + \frac{2\gamma}{g} \) \qquad (0 \le \gamma \le 1).
\label{def:main strip NLIE}
\end{equation}
The main strip of the anti-holomorphic part of $A_1$ NLIE \eqref{fba_s NLIE} is the complex conjugate of the above result.

\paragraph{Source terms of orbifold Konishi.}

We describe the source terms of hybrid NLIE for $(\fa_s \,, \fba_s)$ describing the asymptotic orbifold Konishi state at general $(\alpha, g)$. One can check all these results explicitly by using the formulae in Appendix \ref{app:twisteded Wronskian}.

The holomorphic part of $A_1$ NLIE \eqref{fa_s NLIE} consists of the dynamical variables $(\fa_s \,, \fba_s, Y_{s-2|w})$, and the variables $\fa_s \,, \fba_s$ are related to $\fb_s \,, \fbb_s$ by \eqref{fa to fb}. As reviewed in Appendix \ref{sec:NLIE var}, these variables can be expressed by gauge-covariant ones by
\begin{alignat}{9}
\fb_s &= \frac{Q^{[s+1]} }{\olQ^{[1-s]} } \, \frac{T_{1,s-1} }{L^{[s+1]} } \,, &\quad
1+\fb_s &= \frac{Q^{[s-1]} }{\olQ^{[1-s]} } \, \frac{T_{1,s}^+ }{L^{[s+1]} } \,,
\notag \\[1mm]
1+\fbb_s &= \frac{\olQ^{[1-s]} }{Q^{[s-1]} } \, \frac{T_{1,s}^- }{\olL^{[-s-1]} } \,, &\quad
1+Y_{s-2|w} &= \frac{T_{1,s-1}^- \, T_{1,s-1}^+}{T_{2,s-1} \, T_{0,s-1}} \,.
\label{btoQ fbs}
\end{alignat}

Consider the asymptotic orbifold Konishi state and fix the gauge as given in Appendix \ref{app:transfer orbKon}. For this state, neither Q- nor L-functions have singularities around the real axis, and all critical behaviors come from the extra zeroes of T-functions, $T_{1,s-1}$ and $T_{1,s}$\,, inside the main strip \eqref{def:main strip NLIE}.\footnote{Here we choose the gauge as in }
Since the location of extra zeroes is determined by the values of $(\alpha, g)$, the critical lines $\alpha_{\rm cr} (g)$ are defined by 
\begin{equation}
T_{1,s-1} \( -\ig \) = 0 \qquad {\rm or} \qquad T_{1,s} \( -\frac{i(1-\gamma)}{g} \) = 0 \qquad
{\rm at} \ \ \alpha = \alpha_{\rm cr} (g).
\label{critical line Kon}
\end{equation}
The solution to the equations $T_{1,Q} (-\ig)=0$ also defines the critical lines of the mirror TBA for the twisted \AdSxS, and their asymptotic solutions have been studied in \cite{dLvT11}.
The first equation of \eqref{critical line Kon} has $s-1$ solutions and the second has $s$ solutions for $0 < \alpha < \pi$ and at fixed $g$ with $0 < g \lesssim 1$.\footnote{The equation $T_{1,Q} (-\ig)=0$ has more asymptotic solutions for $g \gtrsim 1$, which are called Type II and Type III critical behaviors in \cite{dLvT11}.}
We denote them by $\alpha_{s-1,i} (g), \ \alpha_{s,i} (g, \gamma)$ with the ordering
\begin{gather}
0 < \alpha_{s-1,1} (g) < \frac{\pi}{s-1} < \alpha_{s-1,2} (g) < \frac{2\pi}{s-1} < \dots
< \frac{(s-2)\pi}{s-1} < \alpha_{s-1,s-1} (g) < \pi,
\notag \\[1mm]
0 < \alpha_{s,1} (g, \gamma) < \frac{\pi}{s} < \alpha_{s,2} (g, \gamma) < \frac{2\pi}{s} < \dots
< \frac{(s-1)\pi}{s} < \alpha_{s,s} (g, \gamma) < \pi.
\label{alpha_cr order}
\end{gather}

It is instructive to keep track of the zeroes of $T_{1,Q}$ in detail, as they behave in an interesting way when $\alpha$ is around $\frac{n \pi}{Q}$ for $n \in \bb{Z},\ 1 \le n \le Q-1$.
If $\alpha$ is slightly less than $\frac{n \pi}{Q}$\,, $T_{1,Q}$ has no zeroes around the real axis.
Let $\alpha$ grow larger.
When $\alpha$ reaches $\frac{n \pi}{Q}$\,, then $T_{1,Q}$ acquires a pair of real zeroes at $\pm \infty$.
The pair of zeroes run toward the origin along the real axis as $\alpha$ increases, and collide at the origin.
After the collision, they run along the imaginary axis in the opposite directions towards $\pm i \infty$.
They cross $\pm \ig$ at $\alpha = \alpha_{\rm cr}^{(i)}$.
There are exceptions at $\alpha = 0, \pi$.
In the limit $\alpha \to 0$, a pair of zeroes of $T_{1,Q}$ run to $\pm \infty$ along the real axis.
Nothing happens around $\alpha = \pi$. As for $\alpha \in (\pi, 2\pi)$ the movement of zeroes is symmetric with respect to the flip $\alpha \to \pi - \alpha$.

\begin{figure}[tb]
\begin{center}
\includegraphics[scale=0.6]{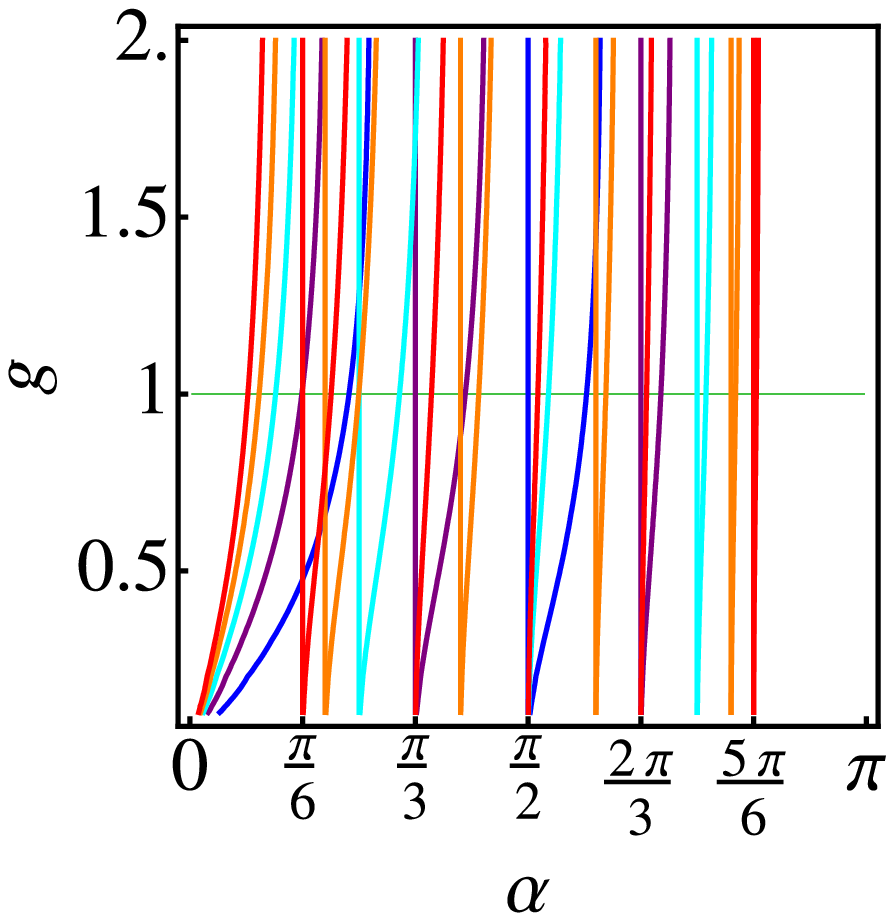}
\hspace{8mm}
\includegraphics[scale=0.6]{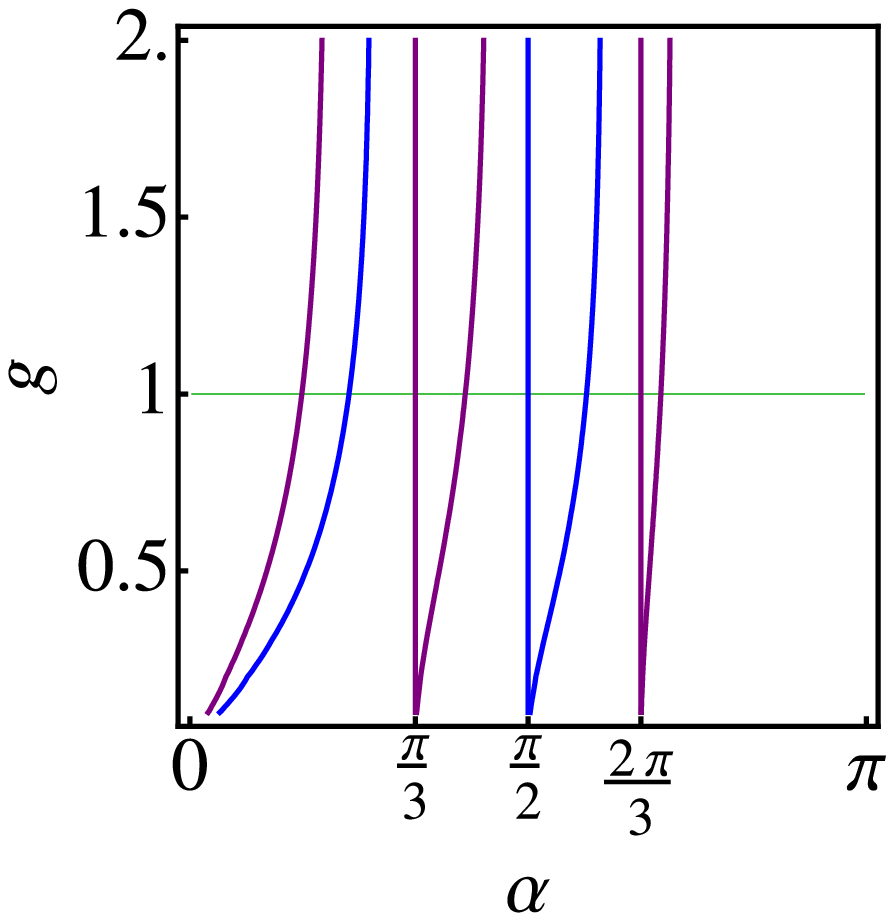}
\caption{Asymptotic phase space of the mirror TBA (Left) and hybrid NLIE (Right) in the horizontal part. We set $s=3$ and $\gamma=0$ in hybrid NLIE. The lines correspond to $\alpha = \frac{n \pi}{Q}$ and the solutions of $T_{1,Q} (-\ig) = 0$ for $Q=2,3,4 \dots$ (TBA) and $Q=2,3$ (NLIE). The phase space of the mirror TBA (Left) should be infinitesimally subdivided if $Q$ is not truncated at $Q=6$.}
\label{fig:phases}
\end{center}
\end{figure}

Let us define the interval
\begin{equation}
I_{s-1} (g) \equiv \bigcup_{n=1}^{s-1} \( \frac{(n-1) \pi}{s-1} \,, \alpha_{s-1,n} (g) \), \quad
I_{s} (g, \gamma) \equiv \bigcup_{n=1}^{s} \( \frac{(n-1) \pi}{s} \,, \alpha_{s,n} (g, \gamma) \).
\label{interval Is}
\end{equation}
Whenever $\alpha$ crosses the boundary of the interval $I_{s-1} (g) \cup I_{s} (g, \gamma)$, the source terms of hybrid NLIE $(J_s \,, \olJ_s)$ change discontinuously.\footnote{Recall that $s=3$ is the minimum choice of hybrid NLIE. In contrast, the phase space $(\alpha, g)$ of the mirror TBA for orbifold Konishi state is classified partially by $\cup_{s=1}^\infty I_s (g)$, which consists of infinitely many segments of the width $\sim \frac{\pi}{s}$ for each $s$.}
The $(J_s \,, \olJ_s)$ at fixed $(\alpha, g)$ are given explicitly as follows.
Start from the source terms for the grounds state \eqref{ground state Js}.
If $\alpha \in I_{s-1} (g)$, add $(j_B \,, \olj_B)$ to $(J_s \,, \olJ_s)$; and then if $\alpha \in I_{s} (g, \gamma)$, add $(j_C \,, \olj_C)$ to $(J_s \,, \olJ_s)$, where $j_B \,, \olj_B \,, j_C \,, \olj_C$ are defined by
\begin{alignat}{9}
j_B (v) &= \sum_j \log S_f \( v - b_j + \frac{i(1-\gamma)}{g} \),
&\quad
\olj_B (v) &= - \sum_j \log S_f \( v - b_j - \frac{i(1-\gamma)}{g} \),
\label{def:jBs} \\[1mm]
j_C (v) &= \sum_j \log S \( v - c_j + \frac{i(1-\gamma)}{g} \),
&\quad
\olj_C (v) &= - \sum_j \log S \( v - c_j - \frac{i(1-\gamma)}{g} \),
\label{def:jCs}
\end{alignat}
where $b_j \,, c_j$ are defined as the zeroes of dynamical variables:
\begin{gather}
1+\fa_s \( b_j - \frac{i(1-\gamma)}{g} \) = 1+\fba_s \( b_j + \frac{i(1-\gamma)}{g} \) = 0,\qquad
b_j \in \cA_{-1+\gamma,1-\gamma} \,,
\label{bj quantize} \\[1mm]
1+Y_{s-2|w} \( c_j - \ig \) = 0,\qquad
c_j \in \cA_{-1,1} \,.
\label{cj quantize}
\end{gather}
All solutions of \eqref{bj quantize}, \eqref{cj quantize} must be summed in \eqref{def:jBs}, \eqref{def:jCs}.
The integral equation for these roots can be obtained by analytic continuation of \eqref{TBA Y1w}-\eqref{fba_s NLIE} as in \cite{AFSu09}, noting that $Y_{s-2|w} (b_j) \propto T_{1,s} (b_j) = 0$.
One can derive the critical lines of \eqref{critical line Kon} from these results, by recalling that $(1+\fa_s), (1+\fba_s)$ are related to $T_{1,s}$\,, and $1+Y_{s-2|w}$ is related to $T_{1,s-1}$\,.
It will turn out in Section \ref{sec:CDT} that each term of \eqref{def:jBs}, \eqref{def:jCs} can be explained by the contour deformation trick of the NLIE \eqref{fa_s NLIE}, where the deformed contour runs through the lower half plane.
Figure \ref{fig:phases} shows the horizontal part of the critical lines in the mirror TBA and hybrid NLIE from the asymptotic analysis.

One remark is needed to evaluate the integrals in TBA and NLIE correctly in a numerical way.
Consider the convolutions $\log (1+\fa_s) \star K_f - \log (1+\fba_s) \star K_f^{[+2-2\gamma]}$ in \eqref{fa_s NLIE}.
If $(1+\fa_s)$ crosses the branch cut of logarithm running the negative real axis, then the integrand changes discontinuously. Suppose there exists $v_d \in \bb{R}$ such that
\begin{equation}
{\rm Im} \[ 1+\fa_s ( v_d ) \] = 0 \ \ {\rm with} \ \ {\rm Re} \[ 1+\fa_s ( v_d ) \] < 0.
\end{equation}
Then we need to integrate $\log (-1) = \pm \pi i$ over $(v_d, \infty)$ or $(-\infty, v_d)$, which provides extra source terms.
As for asymptotic Konishi state, whenever $(1+\fa_s)$ crosses the branch cut of logarithm, then $(1+\fba_s)$ crosses the branch cut at the same point. Thus we get
\begin{align}
\Delta J_s &= - \log \[ S_f (v-v_d) S_f \Big( v-v_d + \frac{2i \, (1-\gamma)}{g} \Big) \] - 2 \pi i,
\\[1mm]
\Delta \olJ_s &= + \log \[ S_f (v-v_d) S_f \Big( v-v_d - \frac{2i \, (1-\gamma)}{g} \Big) \] + 2 \pi i.
\end{align}
The discontinuity of logarithm can in principle happen for the integral with $\log (1+Y_{s-2|w})$.

\section{Contour deformation trick for TBA and NLIE}\label{sec:CDT}

In the last section we studied the ground and orbifold Konishi states in the twisted \AdSxS, in which the hybrid NLIE acquires source terms.
In this section, we turn our attention to the structure of the source term for general states.
It is known that the origin of the source term in the simplified TBA for general states can be explained by both integration of Y-system and contour deformation trick. This is no longer trivially so in hybrid NLIE, as we shall see below.

\subsection{General source terms in the simplified TBA}

Take the simplified TBA for $Y_{1|w}$ as an example, and the following discussion applies to other simplified TBA equations as long as the Y-system exists at that node. We will derive the source terms by integration of Y-system and contour deformation trick.

The explanation by integration of Y-system goes as follows.\footnote{This explanation is also called TBA lemma in the literature.} Consider the logarithmic derivative of Y-system for $Y_{1|w}$
\begin{equation}
\dl \[ Y_{1|w}^- \, Y_{1|w}^+ \] = \dl \[ \(1+Y_{2|w}\) \( \frac{1-\frac{1}{Y_-}}{1-\frac{1}{Y_+}} \) \], \qquad
dl f (v) \equiv \frac{\partial}{\partial v} \log f(v).
\label{dl Ysys 1w}
\end{equation}
Suppose $Y_{1|w} (v)$ has a set of single zeroes $r_j$ inside the strip $\cA_{-1,1}$\,.
If we take the convolution of \eqref{dl Ysys 1w} with $s_K$\,, the left hand side becomes
\begin{equation}
\int_{\bb{R}} dt \; \frac{\partial}{\partial t} \log \[ Y_{1|w} (t^-) Y_{1|w} (t^+) \] s_K (v-t)
= \dl Y_{1|w} (v) + 2\pi i \sum_j s_K \( v-r_j-\ig \).
\end{equation}
Here all solutions of $Y_{1|w} (r_j) = 0, \ r_j \in \cA_{-1,1}$ must be summed.
If we integrate both sides with respect to $v$, we obtain the simplified TBA equation \eqref{TBA Y1w} with\footnote{Note that $\log \frac{1-\frac{1}{Y_-^{[-0]} }}{1-\frac{1}{Y_+^{[+0]} }} \star s_K = \log \frac{1-\frac{1}{Y_-}}{1-\frac{1}{Y_+}} \hstar s_K$ owing to $Y_- (v-i0) = Y_+ (v+i0)$ for $v \in (-\infty,-2) \cup (+2,+\infty)$.}
\begin{equation}
V_{1|w} = c_{1|w} - \sum_j \log S \( v-r_j-\ig \),
\label{V1w TBA lemma}
\end{equation}
where $c_{1|w}$ is an integration constant fixed by the behavior $v \to \pm \infty$, where all Y-functions approach the ground state value.

The explanation by contour deformation trick goes as follows. 
We start from the simplified TBA equation \eqref{TBA Y1w} for the ground state, $V_{1|w} = c_{1|w}$\,. To obtain the TBA equation for excited states, we regard the contour of integration in the right hand side of \eqref{TBA Y1w} as running somewhere far below in the complex plane. When we pull the deformed contour back to the real axis, we obtain additional terms by picking up the residues as
\begin{align}
\log Y_{1|w} &= \log (1+Y_{2|w}) \star_{C_{2|w}} s_K
+ \log \frac{1-\frac{1}{Y_-} }{1-\frac{1}{Y_+} } \hstar_{C_y} s_K \,,
\notag \\
&= - V_{1|w} + \log (1+Y_{2|w}) \star s_K
+ \log \frac{1-\frac{1}{Y_-} }{1-\frac{1}{Y_+} } \hstar s_K \,,
\end{align}
where $C_{2|w} \,, C_y$ are the deformed contour for respective convolutions.

Let $\{ \rho_n \}$ be a set of roots $Y_{1|w} (\rho_n) = 0$, where $\rho_{n} \in \cA_{n-1,n}$ for $n \ge 1$ and $\rho_{n} \in \cA_{n,n+1}$ for $n \le -1$.\footnote{There can be multiple roots as well as poles inside the same strip of the complex plane. It is straightforward to generalize the whole argument for such cases.}
From the Y-system \eqref{dl Ysys 1w} it follows that
\begin{equation}
1+Y_{2|w} (\rho_n^\pm) = 0 \quad {\rm or} \quad
1-\frac{1}{Y_- (\rho_n^\pm)} = 0 \quad {\rm or} \quad
1-\frac{1}{Y_+ (\rho_n^\pm)} = \infty, \qquad n \in \bb{Z}_{\neq 0} \,.
\end{equation}
When we straighten the deformed contours of \eqref{TBA Y1w} running through the lower half plane, the source term $V_{1|w}$ becomes
\begin{equation}
V_{1|w} = c_{1|w} + \log S \( v - \rho_1^- \) + \log S \( v - \rho_{-1}^- \).
\label{V1w CDT}
\end{equation}
where the contributions from $\rho_{-n} \ (n \ge 2)$ vanish owing to $S^- S^+ = 1$.
This result agrees perfectly with \eqref{V1w TBA lemma}.

\subsection{General source terms in $A_1$ NLIE}\label{sec:source terms A1NLIE}

\subsubsection{Fourier transform method}\label{sec:Fourier method}

The $A_1$ NLIE was derived from the assumptions that $Q^{[s-2]}, L^{[+s]}$ are analytic in the upper half plane, and $\olQ^{[2-s]}, \olL^{[-s]}$ are analytic in the lower half plane \cite{Suzuki11}. This derivation can be generalized to the case where dynamical variables have zeroes or poles in the complex plane:\footnote{The Fourier transform of logarithmic derivative diverges if these functions have zeroes on the boundary of $\cA_{m,n}$\,, namely on the line $g \, {\rm Im} \, v \in \bb{Z}$. We should regularize this by shifting the zeroes slightly upward or downward.}
\begin{gather}
T_{1,s} (t_{s,n}) = T_{1,s} (t_{s,-n}) = Q (q_n) = \olQ (\olq_n) = L (\ell_n) = \olL (\olell_n) = 0,
\notag \\[1mm]
\pare{ t_{s,n} \,,\, q_n \,,\, \ell_n } \in \cA_{n-1,n} \,, \qquad
\pare{ t_{s,-n} \,,\, \olq_n \,,\, \olell_n } \in \cA_{-n,-n+1} \,, \qquad (n \ge 1).
\label{def:zeroes TLQ}
\end{gather}
In general, these functions can have multiple zeroes or poles in the complex plane. The generalization for such case is straightforward; if they have poles, the logarithmic derivative have the residue with the opposite sign.
For simplicity we do not discuss poles.

The whole derivation is explained in Appendix \ref{app:derive nlie}. Eventually we obtain the derivative of the source terms $J_s$ appearing in the hybrid NLIE \eqref{fa_s NLIE} as
\begin{equation}
J'_s = J'_s \Big|_T + J'_s \Big|_{\olL} + J'_s \Big|_{L} + J'_s \Big|_{\olQ} + J'_s \Big|_{Q} \,,
\label{summary J's}
\end{equation}
where
\begin{alignat}{9}
\frac{J'_s}{2\pi i} \Big|_T &= - K_f (v - t_{s,1}^-) - K_f (v - t_{s,-1}^-)
- s_K (v - t_{s-1,1}^-) - s_K (v - t_{s-1,-1}^-),
\label{Fourier T src} \\[1mm]
\frac{J'_s}{2\pi i} \Big|_{\olL} &= - \sum_{n=1}^\infty \pare{ K_f (v - \olell_{s+n+1}^{[s-1]} ) + s_K (v - \olell_{s+n}^{[s-2]} ) },
\notag \\[1mm]
\frac{J'_s}{2\pi i} \Big|_{L} &= - \sum_{n=1}^\infty \pare{ K_f (v - \ell_{s+n+1}^{[-s-1]} ) + s_K (v - \ell_{s+n}^{[-s]} ) } - \delta (v - \ell_{s+1}^{[-s-1]} ),
\label{Fourier LL src} \\[1mm]
\frac{J'_s}{2\pi i} \Big|_{\olQ} &= \sum_{n=1}^\infty K_1 (v - \olq_{s+n-1}^{[s-2]}),
\notag \\[1mm]
\frac{J'_s}{2\pi i} \Big|_Q &= \sum_{n=1}^\infty K_1 (v - q_{s+n-1}^{[-s]}) - \delta (v - q_{s+1}^{[-s-1]} ).
\label{Fourier QQ src}
\end{alignat}
We can neglect the $\delta$-functions, as they just add a constant after integration.

\subsubsection{Contour deformation trick with Konishi's contour}\label{sec:CDT}

\begin{figure}[tb]
\begin{center}
\includegraphics[scale=0.8]{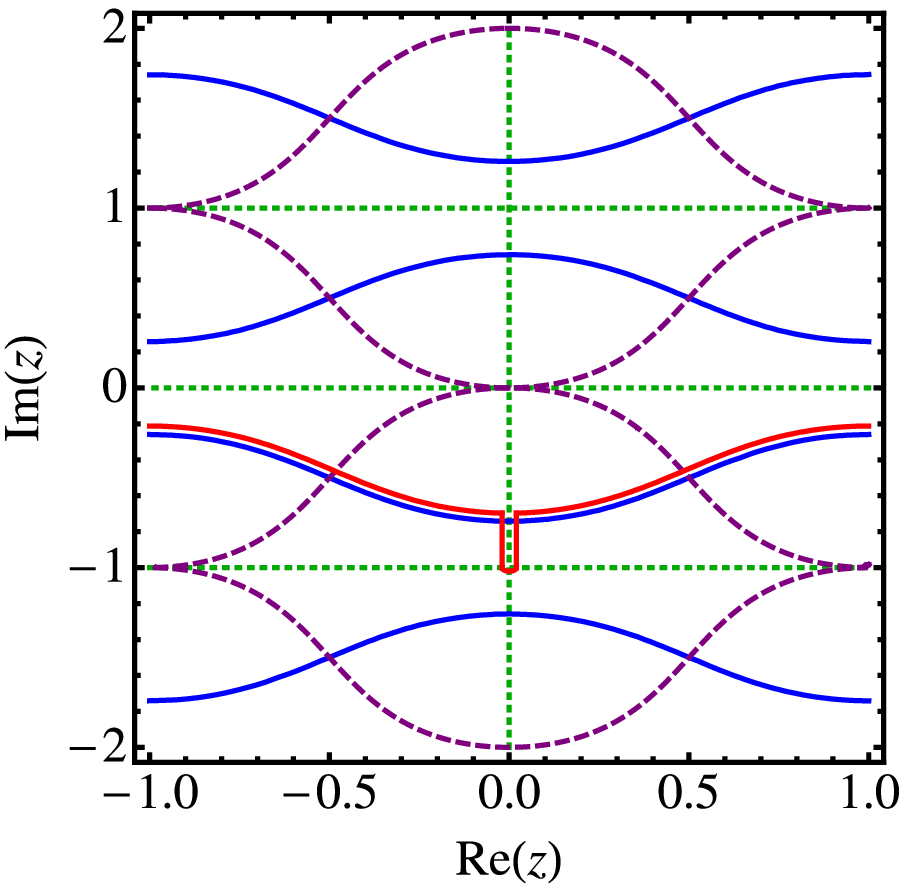}
\hspace{8mm}
\raisebox{8mm}{\includegraphics[scale=1.3]{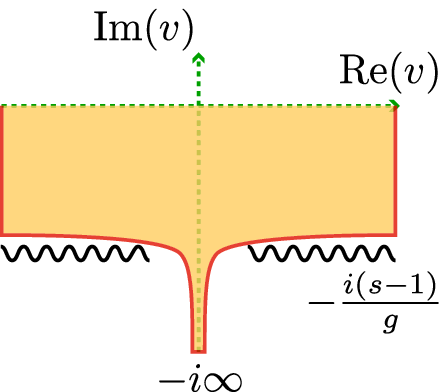}}
\caption{The deformed contour used in the NLIE for $\fb_s$ for the orbifold Konishi state. (Left) the contour in $z$-torus, where the vertical and horizontal axes are normalized by the period of the rapidity torus with moduli $k = -4 g^2/Q^2$, with $Q=s-1$ for $(1+\fb_s), (1+\fbb_s)$ and $Q=s-2$ for $(1+Y_{s-2|w})$. The real line in $z$-torus corresponds to the real axis of the mirror $v$-plane, and the line ${\rm Im} \, z = -1$ corresponds to the real axis of the string $v$-plane. We assumed that there are no singularities like Bethe roots along the string real axis. (Right) the contour in $v$-plane, where the orange region corresponds to the region surrounded by the deformed contour and the mirror real axis.}
\label{fig:deformed contour}
\end{center}
\end{figure}

We start from the $A_1$ NLIE for the ground state with constant source terms $(J_s \,, \olJ_s) = (j_s \,, \olj_s)$.
Then we apply the contour deformation trick to obtain extra source terms, using the same deformed contour as that of the orbifold Konishi state, depicted in Figure \ref{fig:deformed contour}.
For the NLIE of $\fa_s$\,, it runs slightly above the line ${\rm Im} \, v = (1-s+\gamma)/g$, and run down along the imaginary axis. Note that the integrands have branch cut discontinuity along the line ${\rm Im} \, v = (1-s+\gamma)/g$. We take the limit $\gamma \ll 1$ in what follows.

Again we throw the details of computation in Appendix \ref{app:CDT}. After straightening the contour we obtain the following result:
\begin{multline}
J_s^{\rm CDT} = j_s - \log \[ S_f (v - t_{s,1}^-) \, S_f (v - t_{s,-1}^-) \]
- \log \[ S (v - t_{s-1,1}^-) \, S (v - t_{s-1,-1}^-) \]
\\[1mm]
- \log \[ \prod_{j=s+2}^{2s} S_f (v - \olell_j^{[s-1]}) \Big\slash \prod_{j=s+1}^{2s-2} S (v - \olell_j^{[+s]}) \]
+ \log \[ \prod_{j=3}^{s+1} S_f (v - \ell_j^{[-s-1]}) \cdot \prod_{j=3}^{s} S (v - \ell_j^{[-s]}) \]
\\[1mm]
- \log \prod_{j=1}^{s-1} S_1 (v- q_j^{[-s]}) + \log \prod_{j=s}^{2s-2} S_1 (v - \olq_j^{[s-2]}) .
\label{CDT Js}
\end{multline}

\subsubsection{Comparison}

Let us compare the Fourier transform of the derivative of the source terms \eqref{FT J's} (Fourier source terms), with the source terms predicted by the contour deformation trick \eqref{CDT Js} (CDT source terms).
We can make a similar argument for the NLIE of $\fbb_s$\,. Since this is complex conjugate to $\fb_s$\,, we just have to impose the complex-conjugate constraints in addition.

It turns out that there are mismatches in two results. Let us have a closer look for each of the T, L, Q-functions.

\paragraph{T-functions.}

The Fourier source terms \eqref{Fourier T src} agree with the first line of the CDT source terms \eqref{CDT Js}.

\paragraph{L-functions.}

The Fourier source terms \eqref{Fourier LL src} partially agree with the second line of the CDT source terms \eqref{CDT Js}.

The terms with $\{ \olell_m \}$ agree with each other if $\olell_{m \ge 2s-1}$ lie along the imaginary axis in the lower half plane, so that all of them are picked up by the deformed contour.

The terms with $\{ \ell_m \}$ do not agree, because they have the opposite signs. Moreover, the roots $\{ \ell_m \}$ in \eqref{Fourier LL src} lie in the upper half plane, while those in \eqref{CDT Js} lie in the lower half plane.

\paragraph{Q-functions.}

Just like the case of L-functions, The Fourier source terms \eqref{Fourier QQ src} partially agree with the third line of the CDT source terms \eqref{CDT Js}.

If the deformed contour pick up all $\{ \olq_m \}$, then the terms with $\{ \olq_m \}$ perfectly agree with each other.

The terms with $\{ q_m \}$ disagree. The roots $\{ q_{s+n-1}^{[-s]} \}\ (n \ge 2)$ in \eqref{Fourier QQ src} lie in the upper half plane, while those in \eqref{CDT Js} lie in the lower half plane. The corresponding source terms have the opposite signs.
One exception is $q_s^{[-s]}$ in the Fourier source term \eqref{Fourier QQ src}. It lies in the lower half plane, but this term is not present in the CDT source term \eqref{CDT Js}.

\bigskip
The mismatch between two source terms can be explained by different analyticity conditions used in two methods, as summarized in Table \ref{tab:assumptions}. In particular, the extra zeroes of $Q (v)$ at $v \in \cA_{0,s-1}$ and those of $L (v)$ at $v \in \cA_{2,s}$ modify only the CDT source terms.

\begin{table}[tb]
\begin{center}
\begin{tabular}{cl}\hline
Fourier & $Q^{[s-2]} \,, L^{[+s]}$ are meromorphic in the upper half plane.
\\[1mm]
CDT & $Q \,, L^{[+2]}$ are meromorphic in the upper half plane.
\\\hline
\end{tabular}
\caption{Analyticity conditions used in the Fourier transformation method and the contour deformation trick. The complex conjugate conditions for $\olQ, \olL$ are also used. We make no assumptions about $Q \,, L^{[+2]}$ in the lower half plane, $\olQ \,, \olL^{[-2]}$ in the upper half plane.}
\label{tab:assumptions}
\end{center}
\end{table}

Strictly speaking, the T, L, Q-functions may have singularities which can be simultaneously removed by gauge transformation. We forbid such gauge artifacts, and assume that the roots $\{ t_{s,n} \,, \ell_n \,, q_n \,, \olell_n \,, \olq_n \}$ are independent.\footnote{The case of boundstates is exceptional, and further analysis is needed to clarify if the contour deformation trick works as in \cite{AFvT11}.}
In other words, the contour deformation trick with Konishi's contour works fine as long as one can choose a gauge such that all zeroes and poles can be associated to the T-functions rather than the L- and Q-functions.

\subsection{Consistent deformed contour}\label{sec:consistent d-cont}

In the last subsection we have learned that, for states other than the orbifold Konishi, the contour deformation trick with Konishi's contour may not yield the correct source terms of $A_1$ NLIE, as given by the Fourier transform method.
To remedy this problem, we will look for new deformed contours of $A_1$ NLIE.

\bigskip
For the sake of simplicity let us choose the gauge $Q^{\rm I}=\olQ^{\rm I}=1$.
In other words, we will study the analyticity of gauge-invariant quantities,
\begin{equation}
\cT_{1,s} = \frac{T_{1,s} }{Q^{{\rm I} \, [+s]} \, \olQ^{{\rm I} \, [-s]} } \,, \quad
\cL^{[+s]} = \frac{L^{[+s]} }{Q^{{\rm I} [+s]} \, Q^{{\rm I} \, [s-2]} } \,, \quad
\olcL^{[-s]} = \frac{\olL^{[-s]} }{\olQ^{{\rm I} [-s]} \, \olQ^{{\rm I} \, [2-s]} } \,,
\end{equation}
which enables us to rewrite
\begin{equation}
1+\fb_s^{\rm I} = \frac{\cT_{1,s}^+}{\cL^{[s+1]}} \,, \quad
1+\fbb_s^{\rm I} = \frac{\cT_{1,s}^-}{\olcL^{[-s-1]}} \,, \quad
1+Y_{1,s-1} = \frac{\cT_{1,s-1}^- \, \cT_{1,s-1}^+}{\cL^{[+s]} \, \olcL^{[-s]} } \,.
\end{equation}
The zeroes of $\cT , \cL, \olcL$ can be rephrased in terms of analyticity of $\fb_s \,, \fbb_s \,, Y_{1,s-1}$ as,
\begin{alignat}{9}
\cT_{1,s} = 0 \quad &\leftrightarrow &\quad 1+\fb_s^- &= 1+\fbb_s^+ & &= 0,
\notag \\
\cT_{1,s-1} = 0 \quad &\leftrightarrow &\quad 1+Y_{1,s-1}^- &= 1+Y_{1,s-1}^+ & &= 0,
\notag \\
\cL^{[+s]} = 0 \quad &\leftrightarrow &\quad 1+\fb_s^- &= 1+Y_{1,s-1} & &= \infty,
\notag \\
\olcL^{[-s]} = 0 \quad &\leftrightarrow &\quad 1+\fbb_s^+ &= 1+Y_{1,s-1} & &= \infty.
\label{classify TL zeroes}
\end{alignat}
As in Section \ref{sec:source terms A1NLIE}, we consider only the zeroes of $\cT , \cL, \olcL$ and use the notation \eqref{def:zeroes TLQ}. For completeness we also introduce $\cL (\ell_{-n}) = \olcL (\olell_{-n}) = 0$ with $\ell_{-n} \in \cA_{-n,-n+1} \,, \olell_{-n} \in \cA_{n-1,n}$ for $n \ge 1$.

\bigskip
As a warm-up, let us apply the contour deformation trick to $A_1$ NLIE using the contour which encloses all zeroes of $\cT , \cL, \olcL$ in the mirror sheet of complex $v$-plane. 
Just like the contour deformation trick in TBA, we do not pick up the singularities of the kernels.\footnote{The reason for this prescription is not understood.}
Let $*_{\downarrow}$ and $*_{\uparrow}$ be the deformed contours which encloses all zeroes in the lower and upper half plane when pulled backed to the real axis, and $*_\updownarrow \equiv *_\downarrow + *_\uparrow$. We then obtain
\begin{multline}
\log (1+\fb_s) \star_\updownarrow K_f - \log (1+\fbb_s) \star_\updownarrow K_f^{[+2]}
+ \log (1+Y_{1,s-1}) \star_\updownarrow s_K
\\
= - J_s^\updownarrow
+ \log (1+\fb_s) \star K_f - \log (1+\fbb_s) \star K_f^{[+2]}
+ \log (1+Y_{1,s-1}) \star s_K \,,
\end{multline}
with
\begin{multline}
- J_s^\updownarrow = + 2 \log \[ S_f (v-t_{s,1}^-) S_f (v-t_{s,-1}^-) S (v-t_{s-1,1}^-) S (v-t_{s-1,-1}^-) \]
\hspace{30mm}
\\[1mm]
+ \log \[ \prod_{n=1}^\infty S_f (v-\olell_{s+1+n}^{[s-1]} ) S_f (v-\ell_{s+1+n}^{[-s-1]} )
S (v-\olell_{s+n}^{[s-2]}) S (v-\ell_{s+n}^{[-s]})  \]
\\[1mm]
- \log \[ \prod_{k=-\infty, k \neq 0}^{s+1} S_f (v-\ell_k^{[-s-1]}) \, S_f (v-\olell_k^{[s-1]}) \, \cdot
\prod_{k=-\infty, k \neq 0}^s \frac{S (v-\ell_k^{[-s]}) }{S (v-\olell_k^{[+s]}) }\].
\label{Js updown}
\end{multline}
The derivation is discussed in Appendix \ref{app:d-cont var}.

Let us compare the results with the Fourier source terms.
The first line of \eqref{Js updown} involving the zeroes of T-functions is twice as large as \eqref{Fourier T src}, and we should apply the principal value prescription to halve this contribution.
The second line agrees with \eqref{Fourier LL src}, which implies that the third line should be absent.
It is easy to trace the origin of the third line.
For example, $S (v-\ell_k^{[-s]})$ and $S_f (v-\ell_k^{[-s-1]})$ come from the zeroes of $L^{[+s]}$ and $L^{[+s+1]}$ in the lower half plane computed in \eqref{1+Ys down} and \eqref{BK down}, respectively.

Based on this observation, we can specify a deformed contour which is consistent with the Fourier source terms.\footnote{The consistent deformed contour is not necessarily unique, so there is no contradiction with our previous claim on the orbifold Konishi state at weak coupling.}
It turns out that, if we want to apply the contour deformation trick to the consistent deformed contour, we need to study the singularity of integrands first, and classify if they come from T-function or L-functions, following \eqref{classify TL zeroes}.

\begin{figure}[tb]
\begin{center}
\includegraphics[scale=1.0]{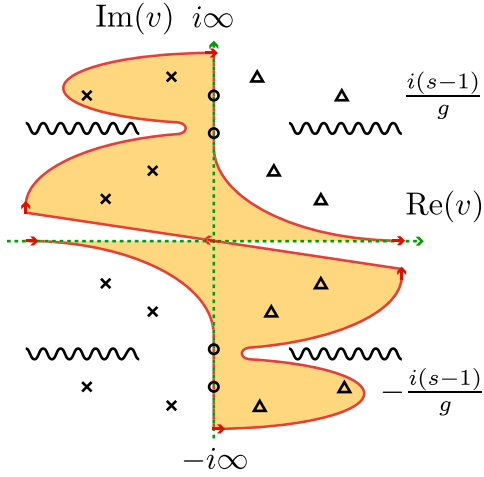}
\caption{The deformed contour for $\fb_s \,, \fbb_s$ in $v$-plane, adjusted to be consistent with the Fourier source term. The symbols $\circ, \times, \triangle$ represent the zeroes or poles of $T, L, \olL$, respectively. The orange region corresponds to the region surrounded by the deformed contour and the mirror real axis.}
\label{fig:deformed symm}
\end{center}
\end{figure}

Let us give one example of the consistent contour by modifying the contours $*_\downarrow \,, *_\uparrow$ to $*_d \,, *_u$\,. For both $*_d$ and $*_u$\,, we make the principal value prescription to the zeroes (or poles) of T-functions. As for $*_d$\,, we neglect the zeroes of $L^{[+s]}$ or $L^{[s+1]}$ in the lower half plane, and as for $*_u$ we neglect the zeroes of $\olL^{[-s]}$ or $\olL^{[-s-1]}$ in the upper half plane.
We join the two contours as shown in Figure \ref{fig:deformed symm}, and denote the corresponding convolution by $*_s = *_d + *_u$\,. We then obtain
\begin{multline}
\log (1+\fb_s) \star_s K_f - \log (1+\fbb_s) \star_s K_f^{[+2]}
+ \log (1+Y_{1,s-1}) \star_s s_K
\\
= - J_s^{\rm cons}
+ \log (1+\fb_s) \star K_f - \log (1+\fbb_s) \star K_f^{[+2]}
+ \log (1+Y_{1,s-1}) \star s_K \,,
\end{multline}
with
\begin{multline}
- J_s^{\rm cons} = \log \frac{S (v-t_{s-1,1}^-)}{S (v-t_{s-1,-1}^+)}
+ \log \[ \prod_{n=1}^\infty \frac{S (v-\ell_{s+n}^{[-s]}) }{S (v-\olell_{s+n}^{[+s]}) } \]
\hspace{50mm}
\\[1mm]
+ \log \[ S_f (v-t_{s,1}^-) S_f (v-t_{s,-1}^-) \]
+ \log \[ \prod_{n=1}^\infty S_f (v-\ell_{s+1+n}^{[-s-1]} ) S_f (v-\olell_{s+1+n}^{[s-1]} ) \].
\label{Js cons}
\end{multline}
The derivation is explained again in Appendix \ref{app:d-cont var}. This result agrees with \eqref{Fourier T src}, \eqref{Fourier LL src}.
Regarding the anti-holomorphic part of $A_1$ NLIE \eqref{fba_s NLIE}, we can construct a consistent deformed contour by taking the complex conjugation.

\bigskip
The source term \eqref{Js cons} depends on the zeroes (or poles) of $T_{1,s-1} \,, T_{1,s}$ in the strip $\cA_{-1,1}$ and the zeroes (or poles) of $L, \olL$ in the upper or lower half planes, $\{ \ell_{s+1} \,, \ell_{s+2} \,, \dots \}, \{ \olell_{s+1} \,, \olell_{s+2} \,, \dots \}$. The latter is related to the poles (or zeroes) of dynamical variables $1+\fb_s^- \,, 1+\fbb_s^+ \,, 1+Y_{1,s-1}$ via \eqref{classify TL zeroes}.
To impose the exact quantization condition on the extra roots lying outside the main strip, we need to analytically continue the NLIE, as mentioned in Section \ref{sec:src A1 NLIE}.
This is a noticeable feature of NLIE compared to the mirror TBA.

\section{Conclusion}\label{sec:conclusion}

In this paper we generalized the hybrid NLIE of \cite{Suzuki11} and applied it to a wider class of states.

First, we studied the ground and the orbifold Konishi states of twisted \AdSxS.
In the mirror TBA, the orbifold Konishi states have infinitely many asymptotic critical lines from $Y_{M|w}$ nodes.
In the hybrid NLIE, the number of critical lines is indeed reduced to a finite number.\footnote{As long as the $\sl(2)$ sector is concerned, this conclusion is expected because the exact truncation method of \cite{Gromov10conf} can be applied without modification.}
The quantization condition for the extra zeroes is written in terms of NLIE variables $(\fa_s \,, \fba_s \,, Y_{1,s-1})$.

Second, we derived the source terms of hybrid NLIE for general states in two ways, Fourier transform method and contour deformation trick.
We constructed the deformed contour which is consistent with the Fourier transform method.

It is interesting to generalize the gauge-invariant NLIE to $A_n$ cases.
The $SU(N)$ principal chiral models contain boundstate spectrum for $N \ge 3$, and its NLIE has been studied in \cite{KL10}. We should be able to reproduce their results by $A_2$ NLIE and contour deformation trick.

While this paper is in preparation, hybrid NLIE of \AdSxS\ made out of $A_1$ and $A_3$ NLIE coupled to the quasi-local formulation of the mirror TBA \cite{BH11b} has appeared in \cite{BH12a}.
We expect that the contour deformation trick will also work to obtain this new NLIE for excited states.

\subsubsection*{Acknowledgements}

The author acknowledges Gleb Arutyunov and Stijn van Tongeren for discussions. This work is supported by the Netherlands Organization for Scientific Research (NWO) under the VICI grant 680-47-602.

\appendix

\section{Notation}\label{app:notation}

We follow the notation of \cite{AFSu09, Suzuki11},
\begin{gather}
x_s (v) = \frac{v}{2} \( 1 + \sqrt{ 1 - \frac{4}{v^2} } \, \),\qquad
x (v) = \frac12 \( v - i \sqrt{4 - v^2} \, \).
\notag \\
\cR_{(\pm)} (v) = \prod_{j=1}^K \frac{x(v) - x^\pm_{s,j}}{\sqrt{x^\pm_{s,j}}} \,,\quad
\cB_{(\pm)} (v) = \prod_{j=1}^K \frac{\frac{1}{x(v)} - x^\pm_{s,j}}{\sqrt{x^\pm_{s,j}}} \,,
\label{def:RBQ}
\end{gather}
together with $f^{[\pm m]} = f ( v \pm \frac{im}{g})$ and $f(v)^\pm = f (v)^{[\pm 1]}$.
The complex rapidity plane are divided into the strips,
\begin{equation}
\cA_{m,n} = \pare{ v \in \bb{C} \ \Big| \ {\rm Im} \, v \in \( \frac{m}{g} \,, \frac{n}{g} \) }.
\label{def:cA}
\end{equation}
We use the following kernels and S-matrices:
\begin{alignat}{9}
s_K (v) & = \frac{1}{2 \pi i} \, \frac{d}{dv} \log S(v)
&\quad S(v) &= -\tanh[ \frac{\pi}{4}(v g - i)] \,,
\notag \\[1mm]
K_Q (v) &= \frac{1}{2\pi i} \, \frac{d}{du} \, \log S_Q (v)
&\quad S_Q (v) &= \frac{v - \frac{iQ}{g}}{u + \frac{i Q}{g}} \,,
\\[1mm]
K_f (v) &= \frac{1}{2\pi i} \, \frac{\partial}{\partial v} \log S_f (v),\qquad
&\quad S_f (v) &= \frac{\Gamma \( \frac{g}{4 \ssp i} \, (v+\frac{2i}{g}) \) \, \Gamma \( - \frac{g \ssp v}{4 \ssp i} \)}{\Gamma \( \ssp \frac{g \ssp v}{4 \ssp i} \) \, \Gamma \( - \frac{g}{4 \ssp i} \, ( v-\frac{2i}{g} ) \)} \,.
\label{def:Kf}
\end{alignat}
One can check the properties $S^+ S^- = 1$ and $S_f^- S_f^+ = S_1$\,.

The convolutions are defined by\footnote{This definition is adapted for Fourier transform and different from the usual convolution in the mirror TBA, {\it e.g.} $F \star K (v) = \int_{-\infty}^\infty dt \, F(t) \, K (t-v)$. Since the kernels $s_K (v)$ is invariant under $v \to -v$, we can still use \eqref{def:star} to write down the simplified TBA for $Y_{M|w}$\,.}
\begin{equation}
F \star K (v) = \int_{-\infty}^\infty dt \, F(t) \, K (v-t),
\qquad
F \hstar K (v) = \int_{-2}^2 dt \, F(t) \, K (v-t).
\label{def:star}
\end{equation}
The logarithmic derivative and its Fourier transform are defined by
\begin{equation}
\dl X (v) \equiv \frac{\partial}{\partial v} \log X(v), \qquad
\hdl X (k) \equiv \int_{-\infty}^{+\infty} dv \; e^{ikv} \frac{\partial}{\partial v} \log X(v).
\label{def:hdl}
\end{equation}
We also use $D_k = e^{k/g}$ and $\hat s_K = 1/(D_k + D_k^{-1})$.
It is useful to keep in mind that the operator $D_k$ shifts the location of zeroes,
\begin{equation}
D_k^n \, e^{ik q} = e^{ik \( q - \frac{in}{g} \)} = e^{ik \, q^{[-n]} } , \qquad
D_k^{-n} \, e^{ik \olq} = e^{ik \, \olq^{[+n]} } .
\end{equation}
Another useful formulae are\footnote{The symbol ${\rm FT}^{-1}$ means the inverse Fourier transform, $\int_{-\infty}^{+\infty} \frac{dk}{2\pi} e^{-ikv}$.}
\begin{align}
{\rm FT}^{-1} \[ \theta (+k) D_k^{+n} \frac{D_k - D_k^{-1}}{D_k + D_k^{-1}} \, e^{ikq} \] &=
- K_f (v - q^{[-n]}) - s_k (v - q^{[1-n]}),
\notag \\[1mm]
{\rm FT}^{-1} \[ \theta (-k) D_k^{-n} \frac{D_k - D_k^{-1}}{D_k + D_k^{-1}} \, e^{ikq} \] &=
+ K_f (v - q^{[+n]}) + s_k (v - q^{[n-1]}).
\label{ift formulae}
\end{align}

The q-number is defined by
\begin{equation}
[s]_q = \frac{q^s - q^{-s}}{q-q^{-1}} \,, \qquad q = e^{i \alpha} \,.
\label{def:q-number}
\end{equation}

\section{Review of NLIE variables}\label{sec:NLIE var}

We briefly review the definition of dynamical variables $(\fa_s \,, \fba_s \,, Y_{1,s-1})$ appearing in $A_1$ NLIE in terms of gauge-covariant variables, the T-, Q- and L-functions \cite{Suzuki11}.
It is convenient to use the gauge-covariant variables when we explain how the source terms of $A_1$ NLIE appear or disappear in accordance with the analyticity of dynamical variables $(\fa_s \,, \fba_s \,, Y_{1,s-1})$.

\subsection{$A_1$ TQ-relations}

It is known that the $A_1$ T-system can be linearized by the $A_1$ TQ-relations \cite{KLWZ96},
\begin{gather}
Q^{[s-2]} \, T_{1,s} - Q^{[+s]} \, T_{1,s-1}^- = \olQ^{[-s]} \, L^{[+s]} \,,
\quad
\olQ^{[2-s]} \, T_{1,s} - \olQ^{[-s]} \, T_{1,s-1}^+ = Q^{[+s]} \, \olL^{[-s]} \,,
\label{def:A1TQ} \\[1mm]
T_{0,s} \, T_{2,s} = L^{[+s+1]} \, \olL^{[-s-1]} \,.
\notag
\end{gather}
As a system of linear difference equations for $Q, \olQ$, these equations have two linearly independent solutions.
We distinguish them by $(Q, \olQ)$ and $(P, \olP)$ if necessary.
We also notice that the equations \eqref{def:A1TQ} are covariant under the gauge transformation of T-system, as discussed in Appendix \ref{app:sym A1TQ}.
In particular, the gauge symmetry becomes manifest if we rewrite \eqref{def:A1TQ} using
\begin{equation}
(Q^{[+s]}, P^{[+s]}, \olQ^{[-s]}, \olP^{[-s]}, L^{[+s]}, \olL^{[-s]})
= (Q_{1,s}^{\rm I}, Q_{1,s}^{\rm II}, \olQ_{1,s}^{\rm I}, \olQ_{1,s}^{\rm II}, L_{1,s} , \olL_{1,s}),
\label{QPL translation}
\end{equation}
as
\begin{gather}
Q_{1,s-1}^{\nu\,-} \, T_{1,s} - Q_{1,s}^\nu \, T_{1,s-1}^- = \olQ_{1,s-1}^{\nu\,-} \, L_{1,s} \,,
\quad
\olQ_{1,s-1}^{\nu\,+} \, T_{1,s} - \olQ_{1,s}^\nu \, T_{1,s-1}^+ = Q_{1,s-1}^{\nu\,+} \, \olL_{1,s} \,,
\label{covariant TQ} \\[1mm]
T_{0,s} \, T_{2,s} = L_{1,s}^+ \, \olL_{1,s}^- \,.
\notag
\end{gather}

The $A_1$ NLIE is written by the gauge-invariant combination of variables in \eqref{covariant TQ} and of the T-system, namely
\begin{equation}
1+\fb_s^\nu = \frac{Q^{\nu\, [s-1]} }{\olQ^{\nu\, [1-s]} } \, \frac{T_{1,s}^+ }{L^{[s+1]} } \,,
\quad
1+\fbb_s^\nu = \frac{\olQ^{\nu \, [1-s]} }{Q^{\nu\, [s-1]} } \, \frac{T_{1,s}^- }{\olL^{[-s-1]} } \,,
\quad
1+Y_{1,s} = 1+Y_{s-1|w} = \frac{T_{1,s}^- \, T_{1,s}^+}{T_{2,s} \, T_{0,s}} \,.
\label{bbY from TLQ}
\end{equation}
For regularization purposes, we define $\fa_s^\nu \,, \fba_s^\nu$ and relate them to $\fb_s^\nu \,, \fbb_s^\nu$ as
\begin{equation}
\fa_s^\nu (v) = \fb_s^\nu \(v-\frac{i \gamma}{g}\), \qquad
\fba_s^\nu (v) = \fbb_s^\nu \(v+\frac{i \gamma}{g}\),
\qquad (0 < \gamma < 1).
\label{fa to fb}
\end{equation}

\subsection{Symmetry in $A_1$ TQ-relations}\label{app:sym A1TQ}

The first line of \eqref{covariant TQ} is invariant under the holomorphic gauge transformation,
\begin{equation}
T_{1,s} \ \to \ g_1^{[+s]} \, g_2^{[-s]} \, T_{1,s} \,,\quad
Q_{1,s}^\nu \ \to \ g_1^{[+s]} \, Q_{1,s}^\nu \,,\quad
\olQ_{1,s}^\nu \ \to \ g_2^{[-s]} \, \olQ_{1,s}^\nu \,,
\label{trans holo}
\end{equation}
provided that the L-functions transform as
\begin{equation}
L_{1,s}^+ \ \to \ g_1^{[s+1]} \, g_1^{[s-1]} \, L_{1,s}^+ \,,\qquad
\olL_{1,s}^- \ \to \ g_2^{[-s+1]} \, g_2^{[-s-1]} \, \olL_{1,s}^- \,.
\label{trans holo LL}
\end{equation}
The TQ-relations are also invariant under the anti-holomorphic transformation,
\begin{equation}
T_{1,s} \ \to \ g_1^{[+s]} \, g_2^{[-s]} \, T_{1,s} \,,\quad
Q_{1,s}^\nu \ \to \ g_2^{[-s]} \, Q_{1,s}^\nu \,,\quad
\olQ_{1,s}^\nu \ \to \ g_1^{[+s]} \, \olQ_{1,s}^\nu \,,
\label{trans anti-holo}
\end{equation}
although it spoils the translational invariance of Q-functions \eqref{QPL translation}.
The combination of two transformations \eqref{trans holo}, \eqref{trans anti-holo} generates a symmetry group larger than the usual gauge transformation of T-system.

The $Y$-functions and the variables $(\fb_s^\nu \,, \fbb_s^\nu)$ are invariant under both transformations:
\begin{equation}
1 + \fb_s^\nu = \frac{Q_{1,s-1}^\nu }{\olQ_{1,s-1}^\nu } \, \frac{T_{1,s}^+ }{L_{1,s}^+ } \,,\qquad
1 + \fbb_s^\nu = \frac{\olQ_{1,s-1}^\nu }{Q_{1,s-1}^\nu } \, \frac{T_{1,s}^- }{L_{1,s}^- } \,.
\label{def:1+bs nu}
\end{equation}
However $(\fb_s^\nu \,, \fbb_s^\nu)$ are not invariant under the frame rotation \cite{GKLV11},
\begin{equation}
\begin{pmatrix}
Q' \\
P'
\end{pmatrix}
= G \begin{pmatrix}
Q \\
P
\end{pmatrix}, \qquad
\begin{pmatrix}
\olQ' \\
\olP'
\end{pmatrix}
= G \begin{pmatrix}
\olQ \\
\olP
\end{pmatrix}, \qquad
G^+ = G^-, \quad
G \in SL (2, \bb{C}).
\end{equation}
This transformation do not change Wronskians $T, L, \olL$, but it acts on the index $\nu$ of $(\fb_s^\nu \,, \fbb_s^\nu)$ in a non-linear way. As a result, the $A_1$ NLIEs before and after the transformation are related in a complicated way.

To write down NLIE we have to specify the frame, {\it i.e.} a particular direction of $\nu$. Due to the nonlinear transformation law of $(\fb_s^\nu \,, \fbb_s^\nu)$ under the frame rotation, it seems to make little sense to consider the $A_1$ NLIE for general $\nu$, or general choice of frame.

\subsection{General solution of $A_1$ TQ-relations}

We look for the most general solution of $A_1$ TQ-relations for given Q-functions, and show that such solution is given by the Wronskian of Q-functions up to a periodic function.

Let us first introduce the differential form as \cite{GKLV11, Volin lec}
\begin{equation}
\bfQ (v) = \sum_{\nu = {\rm I}}^{\rm II} Q^\nu (v) \, {\bf e}^\nu, \qquad
\olbfQ (v) = \sum_{\nu = {\rm I}}^{\rm II} \olQ^{\,\nu} (v) \, {\bf e}^\nu, \qquad
{\bf e}^{\rm I} \wedge {\bf e}^{\rm II} = 1,
\end{equation}
and rewrite the $A_1$ TQ-relations as
\begin{equation}
\bfQ^{[s-2]} \, T_{1,s} - \bfQ^{[+s]} \, T_{1,s-1}^- = \olbfQ^{[-s]} \, L_{1,s} \,,
\quad
\olbfQ^{[2-s]} \, T_{1,s} - \olbfQ^{[-s]} \, T_{1,s-1}^+ = \bfQ^{[+s]} \, \olL_{1,s} \,.
\label{A1TQ form}
\end{equation}
If we apply $\bfQ^{[+s]} \wedge$ and $\wedge \olbfQ^{[-s]}$ to both equations, we obtain
\begin{alignat}{9}
\bfQ^{[+s]} \wedge \bfQ^{[s-2]} \, T_{1,s} &= \bfQ^{[+s]} \wedge \olbfQ^{[-s]} \, L_{1,s} \,,
&\quad
\olbfQ^{[2-s]} \wedge \olbfQ^{[-s]} \, T_{1,s} &= \bfQ^{[+s]} \wedge \olbfQ^{[-s]} \, \olL_{1,s} \,,
\label{A1TQ form2} \\[1mm]
\bfQ^{[s-2]} \wedge \olbfQ^{[-s]} \, T_{1,s} &= \bfQ^{[+s]} \wedge \olbfQ^{[-s]} \, T_{1,s-1}^- \,,
&\quad
\bfQ^{[+s]} \wedge \olbfQ^{[2-s]} \, T_{1,s} &= \bfQ^{[+s]} \wedge \olbfQ^{[-s]} \, T_{1,s-1}^+ \,.\label{A1TQ form3}
\end{alignat}
The equations \eqref{A1TQ form2} are solved by the Ansatz
\begin{equation}
T_{1,s} = A_{1,s} \, \bfQ^{[+s]} \wedge \olbfQ^{[-s]}, \qquad
L_{1,s} = A_{1,s} \, \bfQ^{[+s]} \wedge \bfQ^{[s-2]}, \qquad
\olL_{1,s} = A_{1,s} \, \olbfQ^{[2-s]} \wedge \olbfQ^{[-s]} \,,
\label{TLL general sol}
\end{equation}
and the equations \eqref{A1TQ form2} by
\begin{equation}
A_{1,s} = A_{1,s-1}^- = A_{1,s-1}^+ \,.
\end{equation}
Thus $A_{1,s}$ are periodic functions. This freedom should not be confused with gauge arbitrariness of \eqref{trans holo LL}, because we have already chosen a particular gauge in writing $(\bfQ\,, \olbfQ)$.
These $A$'s cancel out in the combination \eqref{bbY from TLQ}, so without loss of generality we may set them to unity. Then, the general solution \eqref{TLL general sol} becomes the Wronskian as
\begin{gather}
T_{1,s} = \bfQ^{[+s]} \wedge \olbfQ^{[-s]} = {\rm det}
\begin{pmatrix}
Q^{[+s]} & \olQ^{[-s]} \\
P^{[+s]} & \olP^{[-s]} \\
\end{pmatrix},
\label{TLL Wronskian} \\[1mm]
L_{1,s} = \bfQ^{[+s]} \wedge \bfQ^{[s-2]} = {\rm det}
\begin{pmatrix}
Q^{[+s]} & Q^{[s-2]} \\
P^{[+s]} & P^{[s-2]} \\
\end{pmatrix}, \quad
\olL_{1,s} = \olbfQ^{[2-s]} \wedge \olbfQ^{[-s]} = {\rm det}
\begin{pmatrix}
\olQ^{[-s+2]} & \olQ^{[-s]} \\
\olP^{[-s+2]} & \olP^{[-s]}
\end{pmatrix} .
\notag
\end{gather}

\section{Twisted asymptotic data}\label{app:twisteded Wronskian}

Below we summarize the data to solve the mirror TBA and hybrid NLIE for twisted \AdSxS\ in the asymptotic limit.
In particular, we need the twisted transfer matrices written in the form of Wronskian to solve the hybrid NLIE asymptotically.
All T-, L-, Q-functions in this appendix are asymptotic expressions, though we use the same notation as in Appendix \ref{sec:NLIE var}.

\subsection{Generalities}

The twisted transfer matrices of $\alg{su}(2|2)$ symmetry can be constructed by the generating functional called quantum characteristic function \cite{Beisert06b, GL10}. In particular, the quantum characteristic function $D_0$ generates $T_{1,s}$ through
\begin{align}
D_0 &= \( 1 - U_0 T_1 U_0 \) \( 1 - U_0 T_2 U_0 \)^{-1} \( 1 - U_0 T_3 U_0 \)^{-1} \( 1 - U_0 T_4 U_0 \),
\notag \\[1mm]
&\equiv \sum_{s=0}^\infty (-1)^s \, U_0^s \, T_{1,s} (x_0^{[\pm s]}) \, U_0^s \,,
\label{def:D0}
\end{align}
where $U_0$ is the shift operator acting on the mirror rapidity,
\begin{equation}
U^s \ssp f(v) \ssp U^{-s} \equiv f \Big( v + \frac{is}{g} \Big) = f^{[+s]} \,.
\label{def:shift op}
\end{equation}
The $T_n$ are the components of the fundamental transfer matrix, $T_{1,1} = T_1 - T_2 - T_3 + T_4$\,, and they can be written as \cite{AdLST09},\footnote{We introduce $S_0$ since the transfer matrix is defined modulo overall scalar factor.}
\begin{equation}
T_n = S_0 \, \tilde T_n \,,\qquad
S_0 \equiv \prod_{j=1}^\KII \frac{y_j - x^-_0}{y_j - x^+_0}\sqrt{\frac{x^+_0}{x^-_0}} \, \cdot
\prod_{i=1}^\KI \frac{x^+_0-x^+_i}{x_0^+ - x_i^-} \sqrt{\frac{x_i^-}{x_i^+}} \,,
\label{def:tilde Ti}
\end{equation}
with
\begin{alignat}{9}
\tilde T_1 &= \prod_{j=1}^\KII \frac{\nu_j - v - \frac{i}{g}}{\nu_j - v + \frac{i}{g}} \,
\prod_{i=1}^\KI \frac{1 - \frac{1}{x_0^- x_i^+}}{1 - \frac{1}{x_0^- x_i^-}} \sqrt{\frac{x_i^+}{x_i^-}} \,,
&\qquad
\tilde T_2 &= e^{+i \alpha} \, \prod_{j=1}^\KII \frac{\nu_j - v - \frac{i}{g}}{\nu_j - v + \frac{i}{g}} \,
\prod_{k=1}^\KIII \frac{w_k - v + \frac{2i}{g}}{w_k - v} \,,
\notag \\[1mm]
\tilde T_3 &= e^{-i \alpha} \prod_{k=1}^\KIII \frac{w_k - v - \frac{2i}{g}}{w_k - v} \,,
&\qquad
\tilde T_4 &= \prod_{i=1}^\KI \frac{x_0^+ - x_i^-}{x^+_0-x^+_i} \sqrt{\frac{x_i^+}{x_i^-}} \,,
\end{alignat}
where we used $x_0 = x(v),\, x_i = x_s (u_i),\, \nu_j = y_j + 1/y_j$\,, and introduced the twist by\footnote{We rearranged the index $n=1,2,3,4$ from the one used in Section \ref{sec:orbKon}.}
\begin{equation}
T_2 \to e^{i \alpha} \, T_2 \, \qquad T_3 \to e^{-i \alpha} \, T_3 \,.
\label{T2T3 twist}
\end{equation}
By expanding \eqref{def:D0}, we obtain
\begin{align}
T_{1,s} &= \prod_{m=1}^s \( -S_0^{[-s-1+2m]} \) \cdot \[
\tilde \rho_{s+1} - \tilde T_1^{[-s+1]} \, \tilde \rho_s^+
- \tilde \rho_s^- \, \tilde T_4^{[+s-1]} + \tilde T_1^{[-s+1]} \, \tilde \rho_{s-1} \, \tilde T_4^{[+s-1]} \],
\label{T1s by rho} \\[1mm]
\tilde \rho_s &= \prod_{m=1}^{s-1} \tilde T_2^{[-s+2m]}
+ \sum_{k=1}^{s-2} \( \prod_{m=1}^k \tilde T_2^{[-s+2m]} \prod_{n=k+1}^{s-1} \tilde T_3^{[-s+2n]} \)
+ \prod_{n=1}^{s-1} \tilde T_3^{[-s+2n]} \qquad (s \ge 2).
\label{def:rho s}
\end{align}
together with $\tilde \rho_1=1, \tilde \rho_0=0$. 
Note that
\begin{equation}
\prod_{m=1}^s \( - S_0^{[-s-1+2m]} \) =
\prod_{j=1}^\KII \frac{y_j - x^{[-s]}_0}{y_j - x^{[+s]}_0}\sqrt{\frac{x^{[+s]}_0}{x^{[-s]}_0}} \, \cdot
\prod_{m=1}^s \( - \frac{\cR_{(+)}^{[-s+2m]} }{\cR_{(-)}^{[-s+2m]} } \).
\label{S0 product}
\end{equation}

\bigskip
The transfer matrices $T_{1,s}$ \eqref{T1s by rho} can be expressed as the Wronskian of Q-functions in the following way. Let us rewrite $\tilde \rho_{s \ge 1}$ as
\begin{equation}
\tilde \rho_s = \frac{U_3^{[s-1]} }{U_2^{[1-s]} } \sum_{k=0}^{s-1} \varrho^{[-s+1+2k]} \,,
\qquad
\varrho \equiv \frac{U_2}{U_3} \,, \quad 
\tilde T_2 \equiv \frac{U_2^+}{U_2^-} \,, \quad
\tilde T_3 \equiv \frac{U_3^+}{U_3^-} \,,
\label{rho_s as sum}
\end{equation}
and ``differencize" the summation
\begin{equation}
M_\rho^+ - M_\rho^- = \varrho \quad \Rightarrow \quad
M_\rho^{[+s]} - M_\rho^{[-s]} = \sum_{k=0}^{s-1} \varrho^{[-s+1+2k]} \,, \quad
\tilde \rho_s = \frac{U_3^{[s-1]} }{U_2^{[1-s]} } \( M_\rho^{[s]} - M_\rho^{[-s]} \).
\label{differencize rho}
\end{equation}
After a little algebra, \eqref{T1s by rho} becomes
\begin{equation}
T_{1,s} = \prod_{m=1}^s \( -S_0^{[-s-1+2m]} \) \cdot \frac{U_3^{[s-2]} }{U_2^{[2-s]} } \, \sfT_{1,s} \,, \qquad
\sfT_{1,s} =  \ {\rm det}
\begin{pmatrix}
\sfQ^{[+s]} & \olsfQ^{[-s]} \\
\sfP^{[+s]} & \olsfP^{[-s]} \\
\end{pmatrix},
\label{T1s Wronskian}
\end{equation}
where
\begin{alignat}{9}
\sfQ^{[+s]} &= \tilde T_4^{[s-1]} - \tilde T_3^{[s-1]}
= \[ \prod_{i=1}^\KI \frac{x_0^{[+s]} - x_i^-}{x_0^{[+s]}-x^+_i} \sqrt{\frac{x_i^+}{x_i^-}}
- e^{-i \alpha} \prod_{k=1}^\KIII \frac{w_k - v - \frac{i(s+1)}{g}}{w_k - v - \frac{i(s-1)}{g} } \],
\notag \\[1mm]
\olsfQ^{[-s]} &= \tilde T_1^{[1-s]} - \tilde T_2^{[1-s]}
= \prod_{j=1}^\KII \frac{\nu_j - v + \frac{i (s-2)}{g}}{\nu_j - v + \frac{i s}{g}} \[
\prod_{i=1}^\KI \frac{1 - \frac{1}{x_0^{[-s]} x_i^+}}{1 - \frac{1}{x_0^{[-s]} x_i^-}} \sqrt{\frac{x_i^+}{x_i^-}}
- e^{+i \alpha} \, \prod_{k=1}^\KIII \frac{w_k - v + \frac{i(s+1)}{g}}{w_k - v + \frac{i(s-1)}{g} } \],
\notag \\[1mm]
\sfP^{[+s]} &= + \varrho^{[+s]} \, \tilde T_4^{[s-1]} - \sfQ^{[+s]} \, M_\rho^{[+s+1]} \,,
\notag \\[1mm]
\olsfP^{[-s]} &= - \varrho^{[-s]} \, \tilde T_1^{[1-s]} - \olsfQ^{[-s]} \, M_\rho^{[-s-1]} \,.
\label{general QP}
\end{alignat}
It follows that
\begin{alignat}{9}
\sfL^{[+s]} &\equiv {\rm det}
\begin{pmatrix}
\sfQ^{[+s]} & \sfQ^{[s-2]} \\
\sfP^{[+s]} & \sfP^{[s-2]} \\
\end{pmatrix}
& &= \frac{\varrho^{[+s]} \tilde T_3^{[s-1]}}{\tilde T_2^{[s-1]} } \( \tilde T_4^{[s-3]} \sfQ^{[+s]} - \tilde T_2^{[s-1]} \sfQ^{[s-2]} \),
\notag \\[1mm]
\olsfL^{[-s]} &\equiv {\rm det}
\begin{pmatrix}
\olsfQ^{[-s+2]} & \olsfQ^{[-s]} \\
\olsfP^{[-s+2]} & \olsfP^{[-s]}
\end{pmatrix}
& &= \frac{\varrho^{[-s]} \tilde T_2^{[1-s]}}{\tilde T_3^{[1-s]} } \( \tilde T_1^{[3-s]} \olsfQ^{[-s]} - \tilde T_3^{[1-s]} \olsfQ^{[2-s]} \).
\label{general LL}
\end{alignat}

A few remarks are in order.
First, since our twist \eqref{T2T3 twist} affects $T_{1,s}$ only through $\rho_s$\,, the results \eqref{general QP} should formally agree with \cite{GKLT10} modulo gauge transformation.
Second, if one wants to solve a couple of difference equations \eqref{rho_s as sum} explicitly for specific states, it is important to choose a good gauge for T-functions.
Third, for the purpose of getting the asymptotic solution of the hybrid NLIE, we do not have to compute the second set of Q-functions $(\sfP , \olsfP)$. Once we know $\sfT_{1,s} \,, \sfQ, \olsfQ$, we obtain $\sfL, \olsfL$ by the $A_1$ TQ-relations, and they provide sufficient data to construct the gauge-invariant variables $(\fb_s \,, \fbb_s)$.
Fourth, as will be discussed in \eqref{trans holo}, there exists a gauge transformation of T-system which brings the first (or second) set of Q-functions to unity.

\subsection{Transfer matrix for orbifold Konishi}\label{app:transfer orbKon}

Consider the orbifold Konishi state. Since $\KII = \KIII = 0$, it satisfies
\begin{equation}
\tilde \rho_s = \sum_{k=1}^s e^{i \alpha (s+1-2k)}
= \frac{e^{i \alpha s} - e^{-i \alpha s}}{e^{i \alpha} - e^{-i \alpha}} 
= [s]_q \,,
\end{equation}
where $[s]_q$ is the q-number \eqref{def:q-number}. The difference equations \eqref{rho_s as sum}, \eqref{differencize rho} have the solution\footnote{Linear difference equations can be solved by {\it e.g.} Fourier transform.}
\begin{equation}
U_2 = \frac{1}{U_3} = e^{\alpha g v/2}, \qquad
M_\rho = \frac{e^{\alpha g v} - 1}{2 i \sin \alpha} \,, \qquad
(\alpha \neq \pi \bb{Z}).
\label{solUM}
\end{equation}
We added a constant to $M_\rho$ to keep the limit $\alpha \to 0$ non-singular.
The asymptotic Q-functions for the orbifold Konishi state are given by
\begin{alignat}{9}
\sfQ^{[+s]} &= \frac{\cR_{(-)}^{[+s]} }{\cR_{(+)}^{[+s]} } - e^{-i \alpha} 
&\qquad
\olsfQ^{[-s]} &= \frac{\cB_{(+)}^{[-s]} }{\cB_{(-)}^{[-s]} } - e^{+i \alpha} \,,
\notag \\[1mm]
\sfP^{[+s]} &= e^{\alpha (g v + is)} \, \frac{\cR_{(-)}^{[+s]} }{\cR_{(+)}^{[+s]} } - \sfQ^{[+s]} \, M_\rho^{[+s+1]} \,,
&\qquad
\olsfP^{[-s]} &= - e^{\alpha (g v - is)} \, \frac{\cB_{(+)}^{[-s]} }{\cB_{(-)}^{[-s]} } - \olsfQ^{[-s]} \, M_\rho^{[-s-1]} \,,
\label{orbKon QP}
\end{alignat}
and the corresponding $\sfT_{1,s}$ defined in \eqref{T1s Wronskian} is
\begin{equation}
\sfT_{1,s} = e^{\alpha g v} \( [s+1]_q - [s]_q \, \frac{\cR_{(-)}^{[+s]} }{\cR_{(+)}^{[+s]} }
- [s]_q \, \frac{\cB_{(+)}^{[-s]} }{\cB_{(-)}^{[-s]} }
+ [s-1]_q \, \frac{\cR_{(-)}^{[+s]} }{\cR_{(+)}^{[+s]} } \, \frac{\cB_{(+)}^{[-s]} }{\cB_{(-)}^{[-s]} } \).
\label{orbKon T1s}
\end{equation}
We define the L-functions as the solution of the $A_1$ TQ-relations \eqref{def:A1TQ}, which yields
\begin{alignat}{9}
\sfL^{[+s]} &= e^{\alpha g \(v + \frac{i(s-2)}{g}\)} \( 1 + \frac{\cR_{(-)}^{[+s]} }{\cR_{(+)}^{[+s]} } \frac{\cR_{(-)}^{[s-2]} }{\cR_{(+)}^{[s-2]} }
- 2 \cos \alpha \, \frac{\cR_{(-)}^{[s-2]} }{\cR_{(+)}^{[s-2]} } \),
\notag \\[1mm]
\olsfL^{[-s]} &= e^{\alpha g \(v - \frac{i(s-2)}{g}\)} \( 1 + 
\frac{\cB_{(+)}^{[-s]} }{\cB_{(-)}^{[-s]} } \frac{\cB_{(+)}^{[2-s]} }{\cB_{(-)}^{[2-s]} }
- 2 \cos \alpha \, \frac{\cB_{(+)}^{[2-s]} }{\cB_{(-)}^{[2-s]} } \).
\label{Wronskian LL}
\end{alignat}
It also follows that
\begin{equation}
T_{0,s} \, T_{2,s} = T_{1,s}^+ \, T_{1,s}^- - T_{1,s-1} \, T_{1,s+1} 
= L^{[+s+1]} \, \olL^{[-s-1]} = L_{1,s}^+ \, \olL_{1,s}^-.
\end{equation}

Here is a caution for numerical computation. The Wronskian formulae can be numerically unstable at large $|v|$ due to the cancellation of two vectors $(Q, P) \sim (\olQ, \olP)$. To avoid this problem we should use the analytic expression like \eqref{orbKon T1s} instead of the Wronskian form \eqref{T1s Wronskian}. This remark also applies to the L-functions \eqref{Wronskian LL}.

\section{Derivations}

We derive our claims in Sections \ref{sec:source terms A1NLIE} and \ref{sec:consistent d-cont}.

\subsection{Derivation of $A_1$ NLIE with source terms}\label{app:derive nlie}

Below we generalize the derivation of $A_1$ NLIE \cite{Suzuki11} assuming that T, L, Q-functions have zeroes in the complex plane as \eqref{def:zeroes TLQ}, which we repeat here:
\begin{gather}
T_{1,s} (t_{s,n}) = T_{1,s} (t_{s,-n}) = Q (q_n) = \olQ (\olq_n) = L (\ell_n) = \olL (\olell_n) = 0,
\notag \\[1mm]
\pare{ t_{s,n} \,,\, q_n \,,\, \ell_n } \in \cA_{n-1,n} \,, \qquad
\pare{ t_{s,-n} \,,\, \olq_n \,,\, \olell_n } \in \cA_{-n,-n+1} \,, \qquad (n \ge 1).
\label{def:zeroes TLQ}
\end{gather}
The $A_1$ TQ-relations \eqref{covariant TQ} suggest to study the following two variables:
\begin{alignat}{9}
1+\fb_s &= \frac{Q^{[s-1]} \, T_{1,s}^+}{\olQ^{[1-s]} \, L^{[s+1]}} \,, &\quad
\fb_s &= \frac{Q^{[s+1]} \, T_{1,s-1}}{\olQ^{[1-s]} \, L^{[s+1]}} \,,
\notag \\[1mm]
1+\fbb_s &= \frac{\olQ^{[1-s]} \, T_{1,s}^-}{Q^{[s-1]} \, \olL^{[-s-1]}} \,, &\quad
\fbb_s &= \frac{\olQ^{[-s-1]} \, T_{1,s-1}}{Q^{[s-1]} \, \olL^{[-s-1]}} \,.
\label{def:bs su2}
\end{alignat}
Our goal is to deduce the equation of the form $\log \fb_s = \log (1+\fb_s) \star K_f + \dots $ by taking Fourier transform of the logarithmic derivative of these equations. See Appendix \ref{app:notation} for notation.

As a warm-up, consider the T-system at $(1,s-1)$,
\begin{equation}
\hdl \[ T_{1,s-1}^+ T_{1,s-1}^- \] = \hdl \[ (1+Y_{1,s-1}) \, L^{[+s]} \, \olL^{[-s]} \].
\label{Tsys 1,s-1}
\end{equation}
When $T_{1,s-1} (v)$ has zeroes inside the strip $\cA_{-1,1}$\,, we find the relations:\footnote{$T_{1,s-1}$ should not have branch cuts on the real axis, which is asymptotically true for twisted \AdSxS.}
\begin{align}
\hdl T_{1,s-1}^+ &= \int_{\bb{R}+\ig} dv' \, e^{ik (v' - \ig) } \, \partial_{v'} \log T_{1,s-1} (v')
= D_k \pare{ \hdl T_{1,s-1} - 2 \pi i \, e^{ik t_{s-1,1} } } , \quad D_k \equiv e^{+k/g},
\notag \\[1mm]
\hdl T_{1,s-1}^- &= \int_{\bb{R}-\ig} dv' \, e^{ik (v' + \ig) } \, \partial_{v'} \log T_{1,s-1} (v')
= D_k^{-1} \pare{ \hdl T_{1,s-1} + 2 \pi i \, e^{ik t_{s-1,-1} } } .
\label{TBA lemma Ts-1}
\end{align}
The equation \eqref{Tsys 1,s-1} becomes
\begin{equation}
\hdl T_{1,s-1} = \hdl \[ (1+Y_{1,s-1}) \, L^{[+s]} \, \olL^{[-s]} \] \hat s_K
+ 2 \pi i \, \[ D_k \, e^{ik t_{s-1,1} } - D_k^{-1} \, e^{ik t_{s-1,-1} } \] \hat s_K \,.
\label{TpTm 1s-1}
\end{equation}
where $\hat s_K \equiv 1/(D_k + D_k^{-1})$.

\begin{figure}[tb]
\begin{center}
\includegraphics{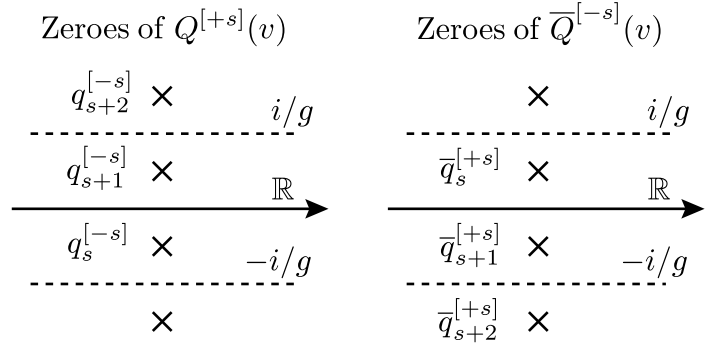}
\caption{Zeroes of $Q^{[+s]} (v)$ and $\olQ^{[+s]} (v)$. Notice that when $Q (v)$ has a zero at $v = q_{s+1} \in \cA_{s,s+1}$ as in \eqref{def:zeroes TLQ}, the shifted function $Q^{[+s]} (v)$ has a zero at $v = q_{s+1}^{[-s]} \in \cA_{0,1}$\,.}
\label{fig:comp zero}
\end{center}
\end{figure}

The relations \eqref{TBA lemma Ts-1} can be generalized to the Q- and L-functions (see Figure \ref{fig:comp zero}):
\begin{alignat}{9}
\hdl Q^{[r+n]} &= D_k^n \hdl Q^{[r]} & &- 2 \pi i \, D_k^{r+n} \sum_{j=1}^n e^{ik q_{r+j}} ,
\notag \\
\hdl Q^{[r-n]} &= D_k^{-n} \hdl Q^{[r]} & &+ 2 \pi i \, D_k^{r-n} \sum_{j=1}^n e^{ik q_{r-n+j}} ,
\notag \\[1mm]
\hdl \olQ^{[-r-n]} &= D_k^{-n} \hdl Q^{[-r]} & &+ 2 \pi i \, D_k^{-r-n} \sum_{j=1}^n e^{ik \olq_{r+j}} ,
\notag \\
\hdl \olQ^{[-r+n]} &= D_k^n \hdl Q^{[-r]} & &- 2 \pi i \, D_k^{-r+n} \sum_{j=1}^n e^{ik \olq_{r-n+j}} ,
\label{hdlQ gen rel}
\end{alignat}
with $r, n \in \bb{Z}_{\ge 1}$\,. By taking the limit $n \to \infty$, we find\footnote{We can derive \eqref{hdlQ cor} also by assuming that $Q$ or $\olQ$ are meromorphic in the upper or lower half plane.}
\begin{alignat}{9}
\hdl Q^{[+s]} &= + 2 \pi i \, D_k^s \sum_{n=1}^\infty e^{ik q_{s+n} } \quad {\rm for} \ {\rm Re} \, k > 0, &\quad
&\( {\rm if} \ \ \lim_{n \to \infty} \ D_k^{-n} \hdl Q^{[r+n]} \to 0 \)
\notag \\[1mm]
\hdl \olQ^{[-s]} &= - 2 \pi i \, D_k^{-s} \sum_{n=1}^\infty e^{ik \olq_{s+n} } \quad {\rm for} \ {\rm Re} \, k < 0, &\quad
&\( {\rm if} \ \ \lim_{n \to \infty} \ D_k^{n} \hdl \olQ^{[-r-n]} \to 0 \).
\label{hdlQ cor}
\end{alignat}

\paragraph{Important lemma.}

In order to derive the NLIE of gauge-invariant variables, it is important to look for a combination of $1+\fb_s \,, 1+\fbb_s$ which do not depend on $T_{1,s}$\,. The answer is
\begin{equation}
\fX_s \equiv \frac{1+\fb_s^-}{1+\fbb_s^+}
= \frac{Q^{[s-2]} Q^{[+s]} \olL^{[-s]} }{\olQ^{[-s]} \olQ^{[2-s]} L^{[+s]} } \,.
\label{def:fXs}
\end{equation}
We then assume that
\begin{align}
Q^{[s-2]} \ {\rm and} \ L^{[+s]} \ &\text{are meromorphic in the upper half plane},
\notag \\
\olQ^{[2-s]} \ {\rm and} \ \olL^{[-s]} \ &\text{are meromorphic in the lower half plane}.
\label{ANZC QL}
\end{align}
These assumptions are realistic, because $Q (v), L (v+\frac{2i}{g})$ do not have branch cuts for ${\rm Im} \, v > 0$ and $s \ge 3$ in our setup.
By applying $\hdl$ on both sides of \eqref{def:fXs}, we obtain
\begin{align}
\hdl \fX_s &= 2 \pi i \!\! \Ress{\rm UHP} \hdl \frac{Q^{[s-2]} Q^{[+s]} }{L^{[+s]} }
+ \hdl \frac{\olL^{[-s]} }{\olQ^{[-s]} \olQ^{[2-s]} } \,,
\qquad ({\rm Re} \, k > 0),
\notag \\[1mm]
\hdl \fX_s &= \hdl \frac{Q^{[s-2]} Q^{[+s]} }{L^{[+s]} } + 2 \pi i \!\! \Ress{\rm LHP} \hdl \frac{\olL^{[-s]} }{\olQ^{[-s]} \olQ^{[2-s]} } \,,
\qquad ({\rm Re} \, k < 0).
\label{hdl fXs applied}
\end{align}
where $\Res_{\,\rm UHP}$ and $\Res_{\,\rm LHP}$ collect the residues in the upper and lower half planes, respectively. By using \eqref{hdl fXs applied} and $\hdl \! f = \theta (+k) \hdl \! f + \theta (-k) \hdl \! f$, we obtain
\begin{align}
\hdl \frac{Q^{[s-2]} Q^{[s]} }{L^{[+s]} } &= + \theta(-k) \hdl \fX_s + 2 \pi i \Res \hdl \fX_s \,,
\notag \\[1mm]
\hdl \frac{\olQ^{[2-s]} \olQ^{[-s]} }{L^{[-s]} } &= - \theta(+k) \hdl \fX_s + 2 \pi i \Res \hdl \fX_s \,,
\label{hdl QQL} \\[1mm]
\Res \hdl \fX_s &\equiv \theta (+k) \! \Ress{\rm UHP} \hdl \frac{Q^{[s-2]} Q^{[+s]} }{L^{[+s]} }
+ \theta (-k) \! \Ress{\rm LHP} \hdl \frac{\olQ^{[-s]} \olQ^{[2-s]} }{\olL^{[-s]} } \,,
\label{def:Res fXs}
\end{align}
The last term can be computed explicitly with the help of \eqref{hdlQ cor} as
\begin{multline}
\Res \hdl \fX_s = \theta (+k) \pare{D_k^s \sum_{n=1}^\infty e^{ik q_{s+n} }
+ D_k^{s-2} \sum_{n=1}^\infty e^{ik q_{s-2+n} } - D_k^s \sum_{n=1}^\infty e^{ik \ell_{s+n} } }
\\[1mm]
+ \theta (-k) \pare{- D_k^{-s} \sum_{n=1}^\infty e^{ik \olq_{s+n} }
- D_k^{-s+2} \sum_{n=1}^\infty e^{ik \olq_{s-2+n} } + D_k^{-s} \sum_{n=1}^\infty e^{ik \olell_{s+n} } } .
\label{app:Res fXs}
\end{multline}

\paragraph{NLIE for $\fb_s$.}

In order to derive the $A_1$ NLIE with source terms, consider $\hdl \fb_s$ in \eqref{def:bs su2},
\begin{align}
\hdl \fb_s &= \hdl Q^{[s+1]} + \hdl T_{1,s-1} - \hdl \olQ^{[1-s]} - \hdl L^{[s+1]} \,,
\notag\\[1mm]
&= D_k^2 \pare{ \hdl Q^{[s-1]} - \hdl L^{[s]} \, \hat s_K }
- \pare{ \hdl \olQ^{[1-s]} - \hdl \olL^{[-s]} \hat s_K } + \hdl (1+Y_{1,s-1}) \, \hat s_K
\notag \\[1mm]
&\qquad - 2 \pi i \, D_k^{s+1} \[ e^{ikq_{s+1} } + e^{ik q_s} - e^{ik \ell_{s+1} } \]
+ 2 \pi i \[ D_k \, e^{ik t_{s-1,1} } - D_k^{-1} \, e^{ik t_{s-1,-1} } \] \hat s_K \,.
\label{hdl fbs comp1}
\end{align}
To rewrite the quantities in the curly brackets, we use $\fX_s$ in \eqref{def:fXs}.
With the help of the formulae \eqref{hdl QQL} and
\begin{align}
\hdl \[ Q^{[s-2]} Q^{[+s]} \] &= \( D_k + D_k^{-1} \) \hdl Q^{[s-1]}
+ 2 \pi i \[ D_k^{s-2} \, e^{ik q_{s-1}} - D_k^s \, e^{ik q_s} \],
\notag \\[1mm]
\hdl \[ \olQ^{[2-s]} \olQ^{[-s]} \] &= \( D_k + D_k^{-1} \) \hdl \olQ^{[1-s]}
- 2 \pi i \[ D_k^{2-s} \, e^{ik \olq_{s-1}} - D_k^{-s} \, e^{ik \olq_s} \],
\label{hres QQ}
\end{align}
we obtain
\begin{multline}
\hdl \fb_s = \pare{ D_k^2 \, \theta(-k) + \theta(k) } \hat s_K \hdl \fX_s
+ \hdl (1+Y_{1,s-1}) \, \hat s_K
+ 2 \pi i \, (D_k^2 - 1) \, \hat s_K \Res \hdl \fX_s
\\[1mm]
+ 2 \pi i \[ - D_k^s \, e^{ik q_{s-1}} - D_k^s \, e^{ik q_s}
- D_k^{2-s} \, e^{ik \olq_{s-1}} + D_k^{-s} \, e^{ik \olq_s} \] \hat s_K
\\[1mm]
- 2 \pi i \, D_k^{s+1} \[ e^{ikq_{s+1} } - e^{ik \ell_{s+1} } \]
+ 2 \pi i \[ D_k \, e^{ik t_{s-1,1} } - D_k^{-1} \, e^{ik t_{s-1,-1} } \] \hat s_K \,.
\label{hdl fbs comp3}
\end{multline}
Since we want an equation of the form $\hdl \fb_s = \hdl (1+\fb_s) \hat K_f + \dots$, we rewrite $\hdl \fX_s$ as
\begin{align}
\hdl \fX_s &= D_k^{-1} \hdl (1+\fb_s) - D_k \hdl (1+\fbb_s) + 2 \pi i \Res \hdl \frac{1+\fb_s^- }{1+\fbb_s^+ } \,,
\label{hres 1+b12} \\[1mm]
\Res \hdl \frac{1+\fb_s^- }{1+\fbb_s^+ } &=
e^{ik t_{s,1} } + e^{ik q_{s-1}^{[-s+2]} } - e^{ik \olq_s^{[+s]} } - e^{ik \ell_{s+1}^{[-s]} }
+ e^{ik t_{s,-1} } + e^{ik \olq_{s-1}^{[s-2]} } - e^{ik q_s^{[-s]} } - e^{ik \olell_{s+1}^{[+s]} } ,
\notag
\end{align}
The last line is the collection of the residues of $\hdl (1+\fb_s)$ inside $\cA_{-1,0}$ and $\hdl (1+\fbb_s)$ inside $\cA_{0,1}$ with appropriate shift.

In summary, Fourier transform of the derivative of $A_1$ NLIE with the source term is
\begin{equation}
\hdl \fb_s = - {\rm FT} \, (J'_s) + \hdl (1+\fb_s) \hat K_f - \hdl (1+\fbb_s) \hat K_f^{[+2]}
+ \hdl (1+Y_{1,s-1}) \, \hat s_K \,,
\end{equation}
where $\hat K_f = \pare{ D_k \, \theta(-k) + D_k^{-1} \theta(k) } \hat s_K$ is the Fourier transform of the kernel $K_f$\,, and 
\begin{multline}
- \frac{{\rm FT} \, (J'_s) }{2\pi i} = D_k \, \hat K_f \Res \hdl \frac{1+\fb_s^- }{1+\fbb_s^+ }
+ (D_k^2 - 1) \, \hat s_K \Res \hdl \fX_s
\\[1mm]
+ \[ - D_k^s \, e^{ik q_{s-1}} - D_k^s \, e^{ik q_s}
- D_k^{2-s} \, e^{ik \olq_{s-1}} + D_k^{-s} \, e^{ik \olq_s} \] \hat s_K
\\[1mm]
- D_k^{s+1} \[ e^{ikq_{s+1} } - e^{ik \ell_{s+1} } \]
+ \[ D_k \, e^{ik t_{s-1,1} } - D_k^{-1} \, e^{ik t_{s-1,-1} } \] \hat s_K \,.
\label{FT J's}
\end{multline}
Here $\Res \hdl \fX_s$ is given in \eqref{app:Res fXs}, and it consists of infinitely many terms.
To obtain \eqref{fa_s NLIE}, we have to apply the inverse Fourier transform and integrate with respect to $v$.\footnote{The formulae \eqref{ift formulae} are useful for this computation.}
The inverse Fourier transform of \eqref{FT J's} is remarkably simple and given by \eqref{summary J's}.
The integration constants can be fixed by consideration of the limit $v \to \pm \infty$.

\paragraph{Case of orbifold Konishi state.}

Let us check if the above results are consistent with the source terms of $A_1$ NLIE for orbifold Konishi state discussed in Section \ref{sec:src A1 NLIE}. As for the asymptotic orbifold Konishi state, $Q^{[s-2]}, L^{[+s]}$ are analytic in the upper half plane and $\olQ^{[2-s]}, \olL^{[-s]}$ are analytic in the lower half plane. We have to take care of the extra zeroes of T-functions only.

Since the $A_1$ NLIE is written in terms of $(\fa_s \,, \fba_s) = (\fb_s^{[-\gamma]}, \fbb_s^{[+\gamma]})$ we have to modify slightly the derivation. In \eqref{hdl fbs comp1} we applied $\hdl$ to the definition of $\fb_s$\,. If we use $\fa_s = \fb_s^{[-\gamma]}$, we obtain
\begin{equation}
\hdl \fa_s = \hdl \fb_s^{[-\gamma]} = D_k^{-\gamma} \[ \hdl \fb_s + 2 \pi i \, e^{i k t_{s-1,-\gamma}} \]
\end{equation}
Actually we may neglect the residue term. After the inverse Fourier transform, it becomes a $\delta$-function, whose integration is just a constant. There is another reason why we do not have to take care of the extra zeroes of $T_{1,s-1}$\,: the rapidity of $Y_{s-1|w}$ in \eqref{fa_s NLIE}, \eqref{fba_s NLIE} is not shifted at all.

An important modification occurs at the equation \eqref{hres 1+b12}, which changes as
\begin{equation}
\hdl \fX_s \equiv D_k^{-1+\gamma} \hdl (1+\fa_s) - D_k^{1-\gamma} \hdl (1+\fba_s) + 2 \pi i \Res \hdl \frac{1+\fa_s^{[-1+\gamma]} }{1+\fba_s^{[+1-\gamma]} } \,,
\end{equation}
Now the last term is the collection of the residues of $\hdl (1+\fa_s)$ inside $\cA_{-1+\gamma,0}$ and $\hdl (1+\fba_s)$ inside $\cA_{0,1-\gamma}$ with appropriate shift.
Since both $(1+\fa_s^{[-1+\gamma]})$ and $(1+\fba_s^{[+1-\gamma]})$ are proportional to $T_{1,s}$\,, this means that the extra zeroes of $T_{1,s}$ inside the strip $\cA_{-1+\gamma,1-\gamma}$ contribute to the source term \eqref{FT J's}. The rest of the derivation goes without any change.

One can see that this conclusion is consistent with the critical behavior observed in \eqref{bj quantize}, \eqref{cj quantize}.

\subsection{Contour deformation for $A_1$ NLIE}\label{app:CDT}

We discuss how to obtain extra source terms in $A_1$ NLIE by applying the contour deformation trick to various deformed contours. When we straighten the deformed contour of the NLIE in the presence of extra zeroes \eqref{def:zeroes TLQ}, we obtain extra terms by collecting the residues.
To simplify the discussion we remove the regulator $\gamma$ by taking the limit $\gamma \ll 1$.

The holomorphic part of $A_1$ NLIE for the ground state ($J_s = j_s$) takes the form
\begin{equation}
\log \fb_s = -J_s + \log (1+\fb_s) \star K_f - \log (1+\fbb_s) \star K_f^{[+2-0]} + \log (1+Y_{s-2|w}) \star s_K \,,
\label{NLIE fb Js}
\end{equation}
where the variables in the right hand side are defined by \eqref{bbY from TLQ}.

\subsubsection{Deformed contour of orbifold Konishi state}\label{app:d-cont Konishi}

For general asymptotic states, $Q, L^{[+2]}$ have no branch cuts in the upper half plane, $\olQ, \olL^{[-2]}$ have no branch cuts in the lower half plane, excluding the real axis. Thus, we can pull the integration contour of $(1+\fb_s), (1+\fbb_s)$ up to ${\rm Im} \, v = \pm (s-1)/g$ and that of $(1+Y_{s-2|w})$ up to $\pm (s-2)/g$. Around the imaginary axis we can further deform them toward $\pm \infty$.

Let $*_K$ be the convolution using Konishi's deformed contour depicted in Figure \ref{fig:deformed contour}.
This contour can pick up all zeroes of T, L, Q-functions inside the strip $\cA_{-s+1, 0}$ or $\cA_{-s+2, 0}$\,. Recalling our notation \eqref{def:zeroes TLQ}, we find\footnote{Use $S_f (v^{[+2]} - t) = S_f (v - t^{[-2]})$ to compute the extra terms from $\log (1+\fbb_s) \star K_f^{[+2]}$.}
\begin{align}
\log (1+\fb_s) \star_K K_f &\to + \log \[ \frac{\ds S_f (v - t_{s,1}^-) \prod_{j=1}^{s-2} S_f (v - t_{s,-j}^-)
\prod_{j=1}^{s-1} S_f (v- q_j^{[1-s]}) }{\ds \prod_{j=s}^{2s-2} S_f (v - \olq_j^{[s-1]})
\prod_{j=3}^{s+1} S_f (v - \ell_j^{[-s-1]}) } \],
\notag \\[1mm]
- \log (1+\fbb_s) \star_K K_f^{[+2]} &\to - \log \[ \frac{\ds \prod_{j=2}^{s} S_f (v - t_{s,-j}^-)
\prod_{j=s}^{2s-2} S_f (v - \olq_j^{[s-3]}) }{\ds \prod_{j=1}^{s-1} S_f (v- q_j^{[-s-1]})
\prod_{j=s+2}^{2s} S_f (v - \olell_j^{[s-1]}) } \],
\notag \\[1mm]
\log (1+Y_{1,s-1}) \star_K s_K &\to + \log \[ \frac{\ds \prod_{j=2}^{s-1} S (v - t_{s-1,-j}^+) \cdot 
S (v - t_{s-1,1}^-) \prod_{j=1}^{s-3} S (v - t_{s-1,-j}^-) }{\ds \prod_{j=3}^{s} S (v - \ell_j^{[-s]})
\prod_{j=s+1}^{2s-2} S (v - \olell_j^{[+s]}) } \],
\end{align}
We assume that all roots $t_{s,-n} (n \ge 1)$ lie along the imaginary axis, as they do for the orbifold Konishi state at weak coupling. Since the deformed contour pick up the corresponding residues, we can replace the upper bound of the product of S-matrices with $t_{s,-n} \,, t_{s-1,-n}$ by $\infty$.

After straightening the contour and using $S^+ S^- = 1$ and $S_f^- S_f^+ = S_1$\,, the source term $J_s$ in \eqref{NLIE fb Js} becomes
\begin{multline}
J_s^{\rm CDT} = j_s - \log \[ S_f (v - t_{s,1}^-) \, S_f (v - t_{s,-1}^-) \]
- \log \[ S (v - t_{s-1,1}^-) \, S (v - t_{s-1,-1}^-) \]
\\[1mm]
- \log \[ \frac{\ds \prod_{j=1}^{s-1} S_1 (v- q_j^{[-s]}) }{\ds \prod_{j=s}^{2s-2} S_1 (v - \olq_j^{[s-2]}) } \cdot 
\frac{\ds \prod_{j=s+2}^{2s} S_f (v - \olell_j^{[s-1]}) }{\ds \prod_{j=3}^{s+1} S_f (v - \ell_j^{[-s-1]}) } \cdot 
\frac{1}{\ds \prod_{j=3}^{s} S (v - \ell_j^{[-s]}) \prod_{j=s+1}^{2s-2} S (v - \olell_j^{[+s]}) } \].
\label{app:CDT Js}
\end{multline}

\subsubsection{Various deformed contours}\label{app:d-cont var}

Below we will derive the results of Section \ref{sec:consistent d-cont}.

The convolutions $*_{\downarrow} \,, *_{\uparrow}$ are defined as the integration with the deformed contour which encloses all zeroes in the lower and upper half plane when pulled backed to the real axis.
Using these deformed contours we obtain the source terms
\begin{align}
&\log (1+Y_{1,s-1}) \star_{\downarrow} s_K
\notag \\
&\quad \to + \log \[
\frac{S (v-t_{s-1,1}^-)}{S (v-t_{s-1,-1}^+)}
\( \prod_{j=1}^\infty \frac{S (v-t_{s-1,-j}^-) \, S (v-t_{s-1,-j}^+) }
{S (v-\olell_{s+j}^{[+s]}) \, S (v-\ell_{-j}^{[-s]} ) } \)
\frac{1}{\prod_{k=1}^s S (v-\ell_k^{[-s]}) }
\],
\label{1+Ys down} \\[1mm]
&\log (1+Y_{1,s-1}) \star_{\uparrow} s_K
\notag \\
&\quad \to - \log \[ 
\frac{S (v-t_{s-1,-1}^+)}{S (v-t_{s-1,1}^-)} 
\( \prod_{j=1}^\infty \frac{S (v-t_{s-1,j}^+) S (v-t_{s-1,j}^-) }
{S (v-\ell_{s+j}^{[-s]}) S (v-\olell_{-j}^{[+s]})} \)
\frac{1}{\prod_{k=1}^s S (v-\olell_k^{[+s]} ) }
\].
\label{1+Ys up}
\end{align}
Similarly, we get
\begin{align}
\log (1+\fb_s) \star_\downarrow K_f &\to + \log \[ S_f (v-t_{s,1}^-)
\( \prod_{j=1}^\infty \frac{S_f (v-t_{s,-j}^-) }{S_f (v-\ell_{-j}^{[-s-1]}) } \)
\frac{1}{\prod_{k=1}^{s+1} S_f (v-\ell_k^{[-s-1]}) }
\],
\label{BK down} \\[1mm]
\log (1+\fb_s) \star_\uparrow K_f &\to - \log \[ \frac{1}{S_f (v-t_{s,1}^- ) }
\(\prod_{j=1}^\infty \frac{S_f (v-t_{s,j}^-) }{S_f (v-\ell_{s+1+j}^{[-s-1]} ) } \)
\],
\label{BK up} \\[1mm]
- \log (1+\fbb_s) \star_\downarrow K_f^{[+2]} &\to - \log \[
\frac{1}{S_f (v-t_{s,-1}^-) }
\( \prod_{j=1}^\infty \frac{S_f (v-t_{s,-j}^- ) }{S_f (v-\olell_{s+1+j}^{[s-1]} ) } \)
\],
\label{BBK down} \\[1mm]
- \log (1+\fbb_s) \star_\uparrow K_f^{[+2]} &\to + \log \[
S_f (v-t_{s,-1}^- )
\(\prod_{j=1}^\infty \frac{S_f (v-t_{s,j}^-) }{S_f (v-\olell_{-j}^{[s-1]}) } \)
\frac{1}{\prod_{k=1}^{s+1} S_f (v-\olell_k^{[s-1]}) }
\].
\label{BBK up}
\end{align}
By adding all of them as $*_\updownarrow = *_\downarrow + *_\uparrow$ and simplifying the result using $S^+ S^- = 1$, we obtain
\begin{multline}
\log (1+\fb_s) \star_\updownarrow K_f - \log (1+\fbb_s) \star_\updownarrow K_f^{[+2]}
+ \log (1+Y_{1,s-1}) \star_\updownarrow s_K
\\[1mm]
\to + 2 \log \[ S_f (v-t_{s,1}^-) S_f (v-t_{s,-1}^-) S (v-t_{s-1,1}^-) S (v-t_{s-1,-1}^-) \]
\hspace{30mm}
\\[1mm]
+ \log \Biggl[
\( \prod_{j=1}^\infty S_f (v-\olell_{s+1+j}^{[s-1]} ) S_f (v-\ell_{s+1+j}^{[-s-1]} )
S (v-\ell_{s+j}^{[-s]}) S (v-\olell_{s+j}^{[s-2]})
\) \ \times
\\[1mm]
\frac{1}{\prod_{j=1}^\infty S_f (v-\ell_{-j}^{[-s-1]}) S_f (v-\olell_{-j}^{[s-1]})
S (v-\olell_{-j}^{[s-2]}) S (v-\ell_{-j}^{[-s]} ) } \ \times
\\[1mm]
\frac{1}{\prod_{k=1}^{s+1} S_f (v-\ell_k^{[-s-1]}) \, S_f (v-\olell_k^{[s-1]}) } \,
\frac{1}{\prod_{k=1}^s S (v-\olell_k^{[s-2]}) S (v-\ell_k^{[-s]}) }
\Biggr],
\end{multline}
which is \eqref{Js updown}.

\bigskip
Another set of contours, $*_d$ and $*_u$\,, are defined as the slight modification of $*_\downarrow$ and $*_\uparrow$\,. For $*_d \,, *_u$ the contribution from the zeroes of T-functions is halved.
The zeroes of $L^{[+s]}$ or $L^{[s+1]}$ in the lower half plane are neglected in $*_d$\,, and the zeroes of $\olL^{[-s]}$ or $\olL^{[-s-1]}$ in the upper half plane are neglected in $*_u$\,.
The contour deformation tricks for \eqref{1+Ys down}-\eqref{BBK up} are now modified as
\begin{align}
\log (1+Y_{1,s-1}) \star_d s_K &\to + \frac12 \log \frac{S (v-t_{s-1,1}^-)}{S (v-t_{s-1,-1}^+)}
- \log \[ \prod_{j=1}^\infty S (v-\olell_{s+j}^{[+s]}) \],
\label{1+Ys down2} \\[1mm]
\log (1+Y_{1,s-1}) \star_u s_K &\to + \frac12 \log \frac{S (v-t_{s-1,1}^-)}{S (v-t_{s-1,-1}^+)}
+ \log \[ \prod_{j=1}^\infty S (v-\ell_{s+j}^{[-s]}) \].
\label{1+Ys up2} \\[1mm]
\log (1+\fb_s) \star_d K_f &\to
+ \frac12 \log \[ S_f (v-t_{s,1}^-) \prod_{j=1}^\infty S_f (v-t_{s,-j}^-) \],
\label{BK down2} \\[1mm]
\log (1+\fb_s) \star_u K_f &\to
+ \frac12 \log \frac{S_f (v-t_{s,1}^- ) }{\prod_{j=1}^\infty S_f (v-t_{s,j}^-) }
+ \log \[ \prod_{j=1}^\infty S_f (v-\ell_{s+1+j}^{[-s-1]} ) \],
\label{BK up2} \\[1mm]
- \log (1+\fbb_s) \star_d K_f^{[+2]} &\to
+ \frac12 \log \frac{S_f (v-t_{s,-1}^-) }{\prod_{j=1}^\infty S_f (v-t_{s,-j}^- ) }
+ \log \[ \prod_{j=1}^\infty S_f (v-\olell_{s+1+j}^{[s-1]} ) \],
\label{BBK down2} \\[1mm]
- \log (1+\fbb_s) \star_u K_f^{[+2]} &\to
+ \frac12 \log \[ S_f (v-t_{s,-1}^- ) \prod_{j=1}^\infty S_f (v-t_{s,j}^-) \],
\label{BBK up2}
\end{align}
By adding all of them and using $*_s = *_d + *_u$\,, we obtain
\begin{multline}
\log (1+\fb_s) \star_s K_f - \log (1+\fbb_s) \star_s K_f^{[+2]} + \log (1+Y_{1,s-1}) \star_s s_K
\\[1mm]
\to + \log \frac{S (v-t_{s-1,1}^-)}{S (v-t_{s-1,-1}^+)}
+ \log \[ \prod_{j=1}^\infty \frac{S (v-\ell_{s+j}^{[-s]}) }{S (v-\olell_{s+j}^{[+s]}) } \]
\hspace{50mm}
\\[1mm]
+ \log \[ S_f (v-t_{s,1}^-) S_f (v-t_{s,-1}^-) \]
+ \log \[ \prod_{j=1}^\infty S_f (v-\ell_{s+1+j}^{[-s-1]} ) S_f (v-\olell_{s+1+j}^{[s-1]} ) \],
\end{multline}
which is \eqref{Js cons}.

\end{document}